# Context-Aware Service Utilisation in the Clouds and Energy Conservation

Saad Liaquat Kiani[1], Ashiq Anjum[2], Nick Antonopoulos[2], Michael Knappmeyer[3], Nigel Baker[1], Richard McClatchey[1]

[1]Faculty of Engineering and Technology, University of the West of England, UK
[2]School of Computing and Mathematics, University of Derby, UK
[3]University of Applied Sciences, Osnabrück, Germany

**Abstract**

Ubiquitous computing environments are characterised by smart, interconnected artefacts embedded in our physical world that are projected to provide useful services to human inhabitants unobtrusively. Mobile devices are becoming the primary tools of human interaction with these embedded artefacts and utilisation of services available in smart computing environments such as clouds. Advancements in capabilities of mobile devices allow a number of user and environment related context consumers to be hosted on these devices. Without a coordinating component, these context consumers and providers are a potential burden on device resources; specifically the effect of uncoordinated computation and communication with cloud-enabled services can negatively impact the battery life. Therefore energy conservation is a major concern in realising the collaboration and utilisation of mobile device based context-aware applications and cloud based services. This paper presents the concept of a context-brokering component to aid in coordination and communication of context information between mobile devices and services deployed in a cloud infrastructure. A prototype context broker is experimentally analysed for effects on energy conservation when accessing and coordinating with cloud services on a smart device, with results signifying reduction in energy consumption.

## 1 Introduction

Ubiquitous computing is characterised by pervasively connected devices that unobtrusively become part of our daily experiences. These computing devices are not just personal computers but include wearable sensors, mobile devices and environmental sensors that project a digital snapshot of the environment, its inhabitants and their activities. Information services based on these interconnected digital artefacts have the potential to enhance our experiences and interactions with the digital world. The role of mobile devices has undergone marked transformation in the last decade, from the initial role of a mobile phone to the modern role of an information acquisition, processing and communication device. These additional roles are enabled by integration of various sensors into the device including GPS receivers, microphones, cameras, magnetometers, RFID readers, proximity and motion detectors in the form of gyroscopes and accelerometers. Combined with the increasing mobility of modern users, availability of various communication technologies (e.g. Bluetooth, GPRS, Wi-Fi) and presence of numerous digital artefacts embedded in the environment, mobile devices are increasingly becoming the primary tools of interaction for inhabitants of the digital world.

In order to enhance our digital experiences and exploit useful services e.g. those deployed by a cloud based service provider,, context-aware applications on our mobile devices can take advantage of availability of information and the ability to interact with surrounding devices and embedded digital artefacts. Context-consuming applications utilise the knowledge created by context providers (services or applications in a cloud) accessible over one of the communication interfaces. Examples of information acquisition carried out by context providers include interfacing with wearable sensors, location and proximity detection and gathering parameters for user activity recognition. Context consumers such as navigation software can use the information gathered by the providers to deliver a useful service to the user. Even with the continuing increase in processing power and computing resources of mobile devices there is still reliance on services external to the device and on-device applications are mostly restricted to context acquisition and consumption. In addition to the sharing of context information between local context providers and consumers, significant context communication takes place between the device-based applications and those deployed in the cloud based network infrastructure. These include network services that depend on information collected at the user device (e.g. location, proximity) and device-based application that depend on context providers hosted in the cloud (e.g. weather service, online calendars, environmental sensors). The traditional user oriented network services, specifically those in the domain of telecom networks, are evolving towards the 'platform/software as a service' model of cloud computing due to reasons of economics, performance and overall efficiency for service providers (e.g. [Vodafone Group, 2010], [Apple Inc., 2011]. This utilisation of information about the user and his environment acquired from the device, network services and artefacts embedded in the environment is the focus of the domain of context-aware clouds.

Technological advancements, social trends and increased utility and adoption of mobile devices mean that context-awareness will be an important element in the next wave of the mobile-device centric cloud services ecosystem. These emerging trends and technologies will provide improved and increased service provisioning mechanisms but the problem remains that the communication and computation of context on smart mobile devices will lead to considerably more energy consumption than that used up in current usage scenarios. This is plausible because network communication and CPU processing in smart phones take up most of the energy resources under typical usage (around 80% on average according to [Anand et al., 2007, p. 1989]. This is a main bottleneck to a wide scale adoption of the cloud-based services in the smart devices, which are being considered as the future platform to consume and produce information. While there are a number of areas where energy conservation approaches can be incorporated (network transmission power, CPU scaling etc.), from the viewpoint of designers of context-aware cloud systems it is imperative that the system design incorporates energy utilisation aspects of software performance. Because of the user centric nature of context-aware applications and the role of mobile devices as the facade of context-aware interaction, this work targets the optimisation of context communication and coordination between the mobile devices and the cloud services in order to conserve energy in mobile devices.

In context-aware systems, context consumers and providers for mobile devices are part of a complex distributed software system (aka cloud system) working towards a common goal of unobtrusively enhancing the user experience and his interaction. The scale, mobility and heterogeneity of devices and users involved in such an environment create challenges in coordination and communication of context across distributed software components in the cloud. The complexity of coordination increases with the increase in inter-dependence between actions of individual components, severity of global constraints faced by the system and the amount of information required by individual components to carry out their tasks. Without the presence of a coordination environment, each device-based consumer and provider has to manage communication and coordination with external services in the cloud individually, resulting in repetition of functionality that incurs a cost in terms of development time and resource usage on the device. While the broader work that forms the basis of the results reported in this paper is targeted towards improving coordination and communication between context consumers and providers, it is the energy consumption related benefits achieved through this approach that are the primary focus of this paper.

The computation power and memory (storage) capacity in mobile devices has continually increased, reducing in cost at the same time, but growth in battery capacity has not followed a similar trend. For example, the specific energy of lithium-ion battery commercialised in early 1990's holds only a small factor more of energy, approx. 100–250 Wh/kg (calculated from specifications provided by Panasonic [2011]), than much older lead-acid batteries (30–40 Wh/kg). Andersson et al. [2006] observe that the development in battery capacity has not followed Moore's law in terms of its evolution. They highlight that while processor capacity continues to follow Moore's law and double every 18 months, battery capacity has only increased 80% in the past 10 years (cf. [Andersson et al., 2006, p. 156]). Because of this limiting factor, even though modern mobile devices can do more, they can do so for only a limited time before the battery runs out of energy. The critical factor of limited energy in mobile devices requires optimised utilisation of resources in aspects of computation, display, interaction and communication with the cloud services. These optimisation mechanisms include on-demand CPU frequency scaling, turning displays off during calls, and variable transmission power for network communications. In addition to these hardware-oriented measures for conserving energy, software lifecycles are designed to optimise the utilisation of resources. Our work builds upon this software-oriented approach where life cycles of device-based and cloud hosted applications and services are managed in a manner that results in energy conservation. We have targeted the applications/services that deal with acquisition, processing and dissemination of contextual information between the smart devices and cloud services and devised a context-brokering system to facilitate the execution of their collective functions.

This paper describes a context brokering system that facilitates coordination and communication of context information (user and environment related context) between context consumers and providers hosted on the device and in the cloud infrastructure. We analyse the effect of introducing a mobile device based Context Broker in our Context Provisioning Architecture on energy consumption on mobile devices by performing a series of experiments. Before describing the experiments it is essential to describe the Context Provisioning Architecture in which the mobile device based Context Broker operates. After discussing the related work in the next section, the functional description of the broker based Context Provisioning Architecture is presented in Section 3. Thereafter, evaluation of the energy conservation aspects is described in Section 5 and 6, followed by the conclusion in the Section 7.

## 2 Background and Related Work

We are able to utilise context in our lives due to a common understanding of the world and oft-recurring situations in it [Winograd, 2001]. The effectiveness of context utilisation is dependent on creating this common understanding. Clarke and Cooper [2000] argue that shared context has to be created to facilitate collaboration in networks, communities and organisations. The established significance of context in human- human interaction is the primary driver for adopting context awareness in human-computer interaction. However, computers cannot take full advantage of context during human-computer interaction due to the limited sensory input available to the machines and a very limited understanding of the world model, as humans perceive it. By increasing the level of contextual understanding in computers, we can increase the richness of communication in human-computer interaction and make it possible to produce more useful computational services [Dey and Abowd, 2000]. This idea that computers can sense and react to stimuli from users' environment is labelled as context-aware computing.

A sizeable number of applications, toolkits and middleware frameworks have been developed to showcase acquisition, processing and distribution of context data in ubiquitous computing environments. Context Toolkit [Dey et al., 1999] offers standard libraries and reusable components to assist in the development of context-aware applications. The Context Broker Architecture (CoBrA) [Chen, 2004] supports context- aware applications in smart spaces. The PACE [Henricksen et al., 2005] middleware offers functionalities for context gathering, context management and context dissemination. The Service-Oriented Context-Aware Middleware (SOCAM) [Gu et al., 2004] project provides architecture for building and rapidly prototyping context-aware mobile services. An example for a centralised middleware approach designed for context-aware mobile applications is the Context-Awareness Sub-Structure (CASS) [Fahy and Clarke, 2004]. These are typical examples of middleware solutions and context-aware systems that have been proposed in the last decade. A common theme in these systems is reliance on a central component or server to collect, manage and distribute context information to interested applications or services and the role of mobile devices (and hosted applications) as more than mere consumers of context is not adequately considered. Moreover, context providing and consuming components are tightly coupled and demonstrations of the systems are limited to a specific knowledge domain, e.g. tourist guides and meeting room scenarios. Recent developments in this domain have attempted at decoupling consumers and providers of context information by employing a context broker, e.g. MobiLife [Kernchen et al., 2006] and Context Casting [Knappmeyer et al., 2009, Moltchanov et al., 2010].

The concept of brokering is formalised as an architectural design pattern, known as the Broker Pattern [Buschmann et al., 2007], that can be used to structure distributed software systems with decoupled components that interact by remote service invocations (cf. Buschmann et al. [2007, p. 99]) or message passing. The use of brokers in information distribution systems is well established e.g. in CORBA [OMG, 2008] and Web Services Brokered Notification [Huang and Gannon, 2006]. In the domain of context-aware computing, broker-based architectures have been developed for collecting, processing and distributing user and environmental context. In contrast to centralised context brokers demonstrated in these approaches, this work extends the concept and utilises a federation of context brokers for dissemination of context information between cloud based distributed consumers and providers.

Despite a number of context-aware systems in existence, energy consumption and conservation on mobile devices involved in context acquisition and utilisation has not been adequately addressed. Earlier systems were developed with the aim of demonstrating various functions of context-aware systems, e.g. context acquisition, management, representation, reasoning, and their prototype nature did not consider energy constraints. Moreover, the role of mobile devices in most context-aware systems has been limited to consuming context information in the form of a single executing application on the device. However, as discussed in the preceding section, the role of mobile devices is becoming more central to our interaction with the digital world and their increased capabilities allow concurrent execution of a number of context consuming and producing applications and services. More energy is utilised on the device in this evolving role and hence adoption of energy conservation techniques in mobile device based context consumers and providers becomes even more important.

In general the area of energy conservation on smart devices has garnered significant interest from re- searchers and equipment manufacturers. Various energy conservation measures are embedded in the design of mobile devices, e.g. dimming of the display screen when not in use, turning off the display during calls, switching off secondary radios (Bluetooth) during low battery conditions. Takeno et al. [2003] studied the patterns of energy consumption in consumer mobile phones along with charging and discharging life cycles of consumer mobile devices. They attempted to predict the lifetime of a Lithium-ion battery considering deterioration in capacity and showed that the charge state and the number of recharges influence the battery capacity during storage. Ravi et al. [2008] have proposed the CABMAN system for battery management on mobile phones that warns the user when it detects that the phone battery can run out before the next charging opportunity is encountered. CABMAN system is based on a context-aware algorithm that monitors device usage, location and battery charge pattern to predict when the next charging opportunity will be available, how much call-time will be required by the user in the interim, and how long the battery will last if the current set of applications continues to execute. Mahmud et al. [2004] have carried out a comprehensive study on energy consumption in multi-interface mobile devices and proposed a model for predicting energy consumption during various types of data exchanges. Gupta and Mohapatra [2007] provide a detailed anatomy of the energy consumption by various components of Wi-Fi based phones and emphasise on adopting an adaptive scanning algorithm, which results in considerable energy savings. Kassinen et al. [2009] investigate how long a mobile peer in a P2P network is able to function in a UMTS or WLAN access network, and how the different parameter settings affect this battery life. They conclude that a realistic example of a useful improvement could be a connectivity management system that switches from UMTS to WLAN when possible, to lengthen the average battery life of mobile peers. Similarly, Rahmati and Zhong [2007] highlight the challenges of energy costs of network interfaces in mobile devices in their field-study and propose to leverage the complementary strength of Wi-Fi and cellular networks by choosing wireless interfaces for data transfers based on network condition estimation. They show that an ideal selection policy can more than double the battery lifetime of a commercial mobile phone, and the improvement varies with data transfer patterns and Wi-Fi availability. Balasubramanian et al. [2009] reiterate the notion that energy on mobile phones is a precious resource and attempt to model relative energy consumption characteristics of 3G, GSM and Wi-Fi. They identify a significant tail energy overhead in 3G and GSM and develop a measurement driven model of energy consumption of network activity for each technology. Based on this model they have also developed TailEnder, a protocol that minimises energy usage by prefetching data while meeting delay-tolerance deadlines specified by users.

Several schemes have been proposed to improve energy utilisation by optimising the device's air interface. Yang [2007] investigates the power saving mechanism of UMTS and proposes an adaptive algorithm to en- hance UMTS energy consumption performance. Another version of this algorithm for bursty traffic patterns is provided in [Yang et al., 2007]. In addition to these optimisations aimed at energy conservation on the device side, conservation on the network side has also been notably addressed e.g. using dynamic network planning to reduce the number of active access devices in the core network when they are under-utilised [Chiaraviglio et al., 2009].

In the specific domain of context-aware systems, it is only recently that attention has been paid towards analysing the impact of context-aware applications on the energy consumption.. Bernal et al. [2010] have proposed a mechanism for reducing the context data publishing from mobile devices by adapting context publishing behaviour according to device conditions (signal strength, sensors status etc.). Similarly Kang et al. [2008] emphasise that the major challenge in providing users with proactive services lies in continuously acquiring user and environment context from sensors due to the imposition of heavy workloads on mobile devices and energy consumption in performing this task. They attempt to address this challenge in their context-monitoring framework for sensor-rich and resource limited mobile environments, titled 'SeeMon'. The energy efficiency in SeeMon is based on optimal selection of the Essential Sensor Set (ESS) that can satisfy a context query. Crk et al. [2008] propose a range of user-interaction-aware mechanisms that utilise a novel approach of monitoring a user's interaction with applications through the capture and classification of mouse events to effect considerable improvements in energy savings and delay reductions of the Wi-Fi network interface. Devlic et al. [2008] emphasise the need to retrieve context information from outside the device via network interfaces and evaluate the use of Bluetooth and Wi-Fi interfaces for this purpose. They conclude that multicasting over Wi-Fi consumes less energy than Bluetooth and that it is more energy efficient to distribute context knowledge to other devices than having each device learn or acquire this information itself. Zhuang et al. [2010] analyse the issue of energy efficient location sensing on smart phones and identify four critical factors that affect energy efficiency of location sensing with GPS sensors. These factors include static use of location sensing mechanisms, absence of power-efficient sensors, lack of sensing cooperation among multiple location based applications and unawareness of the battery state. They present an adaptive location-sensing framework for Android-based smart phones; evaluation results show significant reduction in the GPS usage and improvement in battery life. The efforts discussed in these lines showcase the fairly recent trend recognising the importance of energy conservation in context-aware systems.

The preceding paragraphs have shown that there are various approaches that are being targeted for reduction in energy consumption. However, most of the efforts in the domain of context- awareness deal with particular applications (location sensing, Wi-Fi usage, etc.) whereas this study focuses on emerging scenarios where a comparatively larger number of context consuming and producing applications are deployed in a cloud and consumed by mobile device based applications. With recent advancements in the capabilities of mobile devices and the surrounding digital ecosystem, we can safely expect an increase in the amount of contextual information acquired, requested and published from the mobile devices. In light of this trend we aim to investigate the energy conservation benefits of the brokering approach used in the Context Provisioning Architecture during communication of contextual information between cloud services and mobile devices. The following section provides a description of the Context Provisioning Architecture, including the functional design of the Context Broker.

# 3 Context Provisioning Architecture

The Context Provisioning Architecture is based on the producer (provider)-consumer model in which cloud services take the roles of context providers or context consumers. These basic entities are interconnected by means of a context broker that provides routing, event management, and query resolution and lookup services. The following paragraphs describe these three main components of the architecture.

**Context Consumer** A Context Consumer (CxC) is a component (e.g. a context based application) that uses context data. A CxC can retrieve context information by sending a subscription to the Context Broker (CxB) or a direct on-demand query and context information is delivered when and if it is available.

**Context Provider** The Context Provider (CxP) component provides context information. A CxP gathers data from a collection of sensors, network/cloud services or other relevant sources. A CxP may use various aggregation and reasoning mechanisms to infer context from raw sensor, network or other source data. A CxP provides context data only further to a specific invocation or subscription and is usually specialised in a particular context domain (e.g. location).

**Context Broker** A Context Broker (CxB) is the main coordinating component of the architecture. It works as a facilitator between other architectural components. Primarily the CxB has to control context flow among all attached components, which it achieves by allowing CxCs to subscribe to context information and CxPs to deliver notifications.

A depiction of the core system components described above is presented in Fig. 1, emphasising the complementary provision of synchronous and asynchronous context related communication facilities. A number of useful applications have been developed based on this architecture. Further details on this architecture and industrial trials are described in [Moltchanov et al., 2008, Zafar et al., 2009, Knappmeyer et al., 2011].

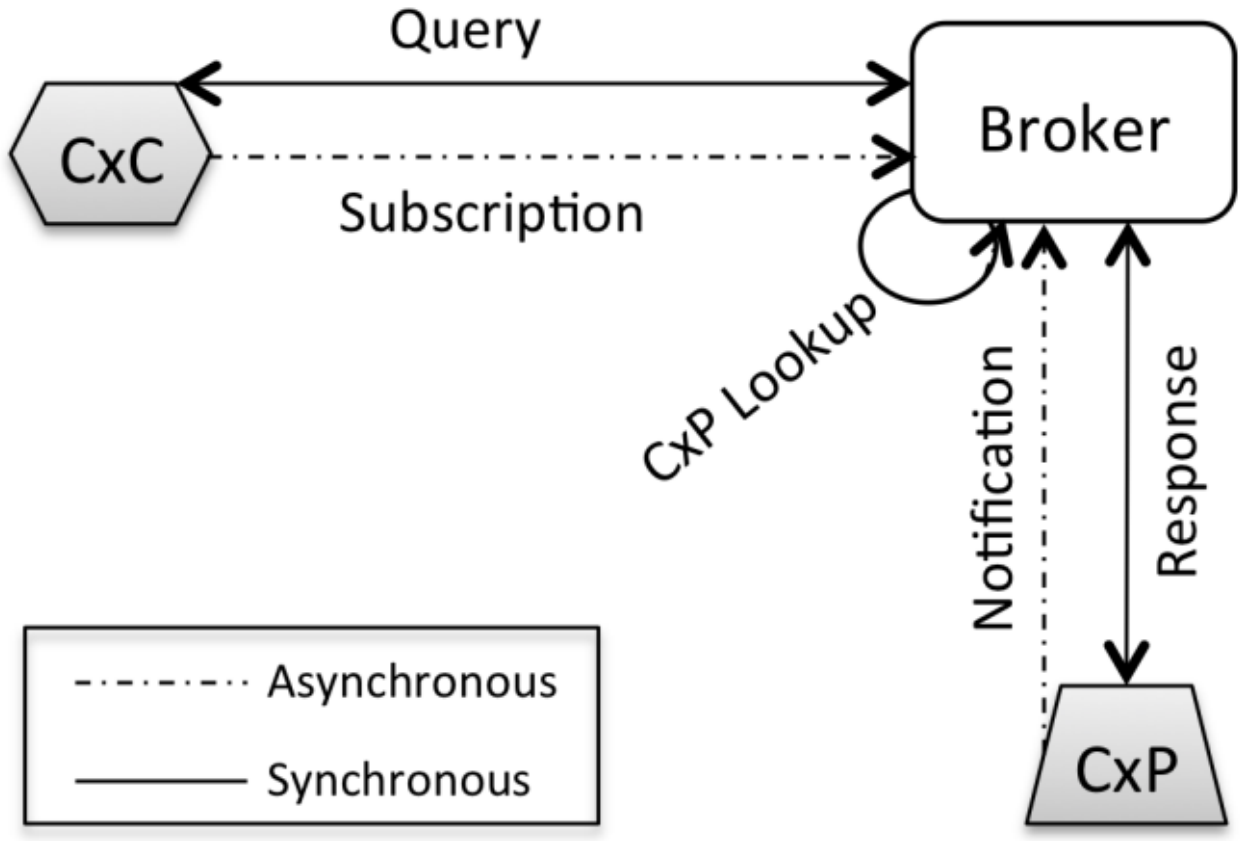

Figure 1: Functional components of the Context Provisioning Architecture.

The following example elaborates the working of the Context Provisioning Architecture. Consider a cloud service hosted by a telecom operator that builds multicast groups based on user location for efficient delivery of multimedia content to its mobile users. Such a service can query user interests and preferences from a user profile provider and location from GPS based location providers being executed on user devices. The type of content may further be customised based on a user's current activity, which can be acquired from an activity context provider being executed partly on the user device and partly on a server deployed in the cloud infrastructure. The activity context provider can use the on-board accelerometers, noise and light sensors, and running applications on the device to estimate the current activity context of the user (e.g. [Choudhury et al., 2008, Fábián et al., 2008]. The provider-consumer interaction is depicted in Fig. 2.

Building on this complex scenario, it can be deduced that a number of context providers and consumers on the user device will be interacting with a context broker on the cloud network (NCxB) for acquisition and provision of context data within the context aware system and beyond (e.g. third party services). If each device level context provider and consumer were to handle broker-bound communication itself, not only the computation cost will increase but also the development cost for new providers and consumers. Given the inherent mobility and somewhat intermittent connectivity of mobile devices, each provider and consumer may have to carry out extra life-cycle management tasks as well. To mitigate the effects of these issues, a device-based broker is utilised in this architecture that works in federation with the context brokers deployed in the cloud infrastructure. This Mobile Context Broker (MCxB) executes on the user mobile device and facilitates the device level context providers and consumers in retrieving and providing context to cloud infrastructure components through a publish/subscribe mechanism.

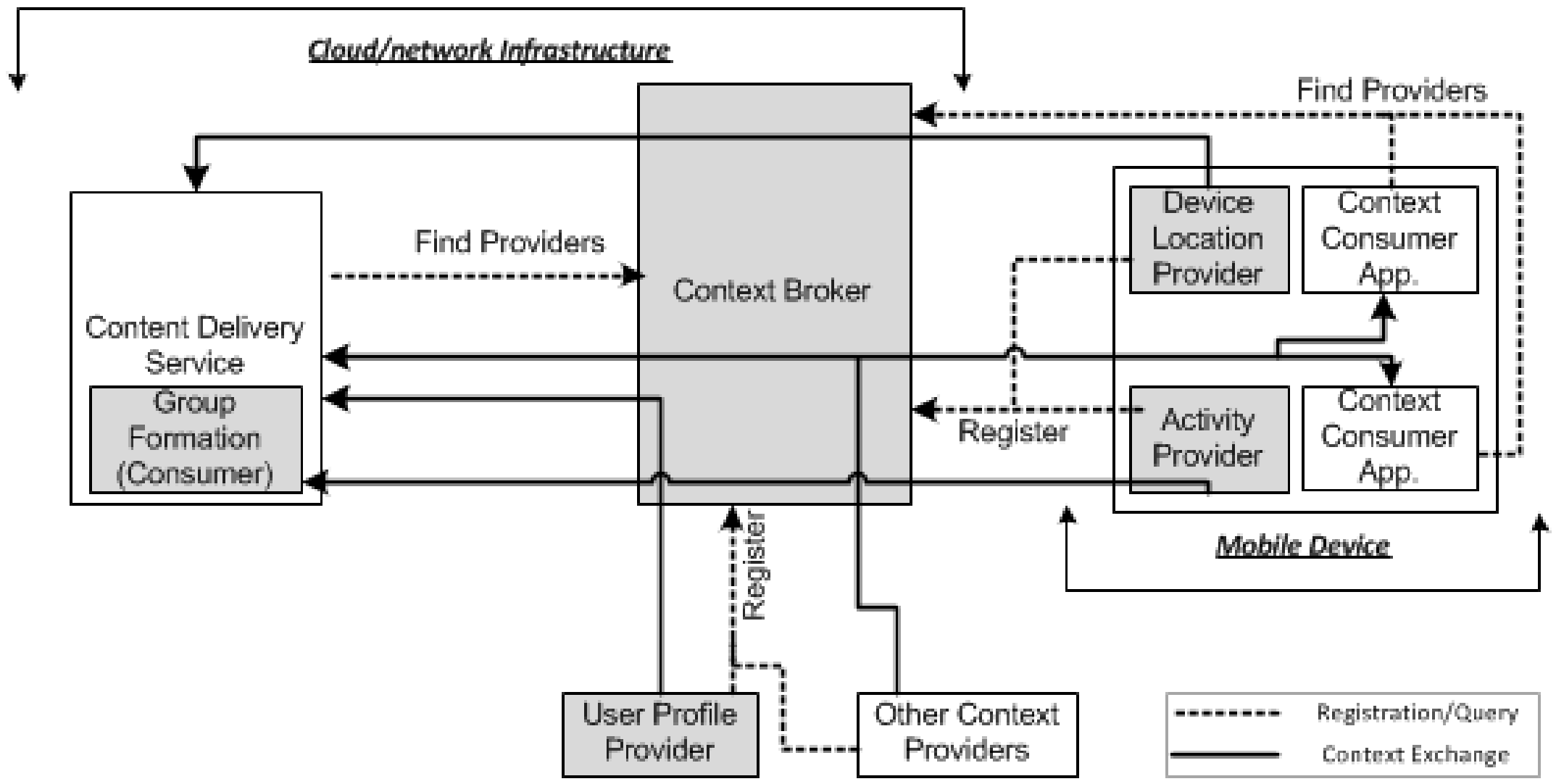

Figure 2: Broker based interaction: demonstrating the use of a context broker to lookup required context providers and consumers.

## 3.1 Mobile Context Broker

The Mobile Context Broker is a software component designed to execute on a mobile device as a background service that brokers context exchange between consumers and providers, hosted both on the device and the cloud infrastructure. Context providers and consumers register their presence and requirements during execution to this broker and do not have to lookup each other individually. Moreover, during periods of disconnected operation, which is still common in mobile devices and networks, the consumers and providers do not have to monitor device connectivity individually; this task is delegated to the MCxB. Polling and waiting for events or context information to become available by consumer components is improved by applying the publish/subscribe communication paradigm and using the broker as an event service that manages notifications and subscriptions. These functions provided by the broker save valuable computation cycles and consequently reduce energy consumption. Figure 3 shows the addition of a device-based broker in the scenario discussed earlier (cp. Fig. 2). The design, functional architecture and the coordination model that enable these advantages of our broker-based architecture are presented in the following paragraphs.

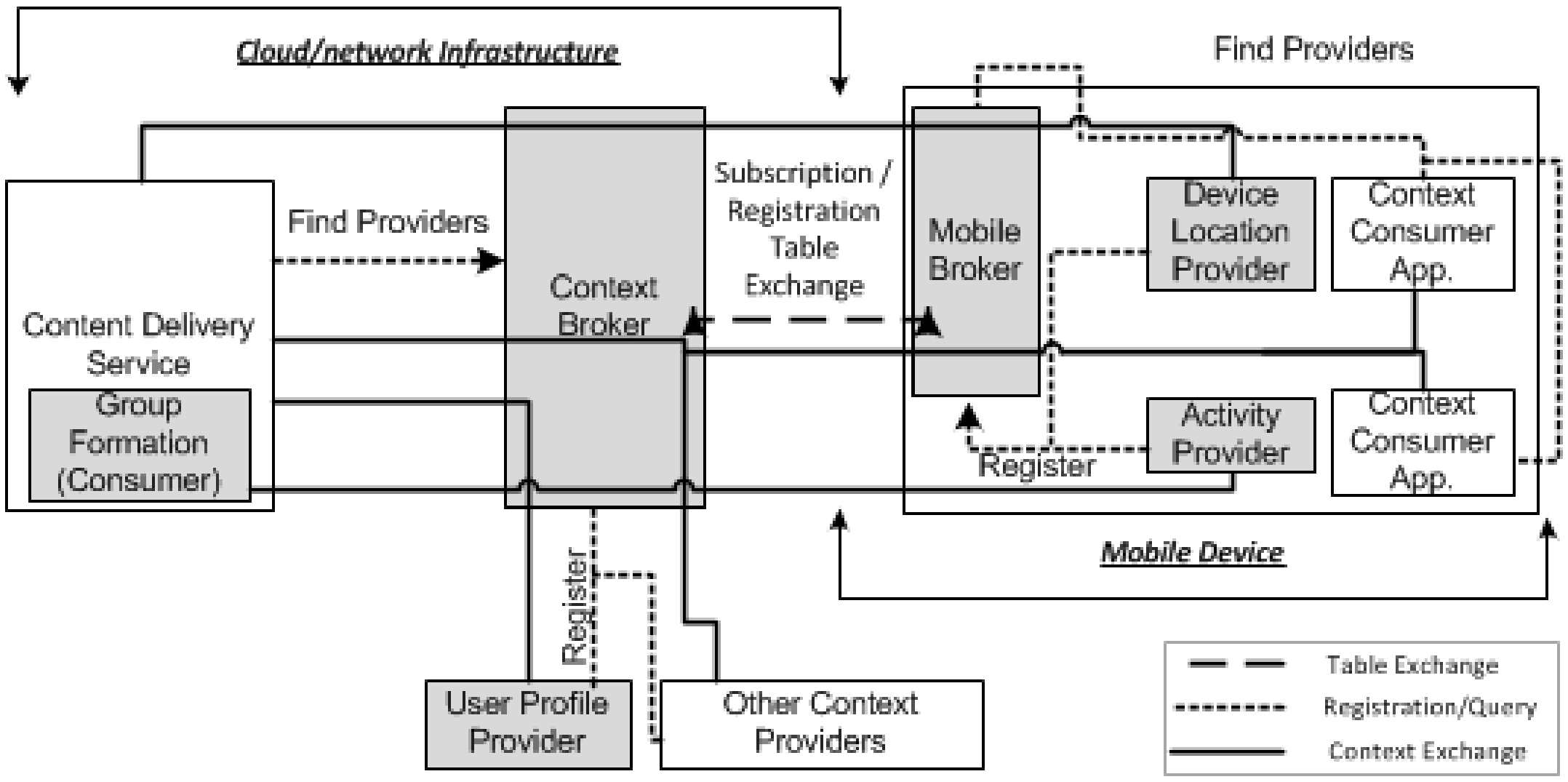

Figure 3: Functional architecture of the Context Provisioning Architecture with mobile broker component.

The design of the broker is based on the set of functions it provides to the context consumers and providers. These functions are listed below:

**Registration and Lookup** Each consumer and provider registers with a broker by specifying its communication end point and the type of context it provides or requires. This function in turn enables the brokering function in which the mobile broker can direct a consumer requesting a particular type of context to the correct provider(s).

**Subscription** A consumer subscribes with the broker specifying the type and instance of context it requires and the duration for which the subscription remains valid. The broker can forward the subscription to the appropriate provider or filter context produced by a provider in order to satisfy the subscription.

**Notification and Dissemination** The broker, on availability of subscription satisfying context, notifies the consumer of the availability or the context is directly communicated to the consumer.

**Caching** The broker can cache recently produced context in order to exploit the principle of locality of reference, as done routinely in internet communications, to improve overall performance.

**Querying** The broker provides a query resolution service via which context consumers can request the broker to fetch context from the providers. The querying function is equivalent to a one-off subscription, valid only once irrespective of whether it results in meaningful response from the provider or unavailability of information.

**Bulk Query and Response Management** The Mobile Context Broker can operate in *bulk query mode* for low priority queries in which it forwards queries and responses towards the cloud in bulk. This store and forward bulk mode is useful not only in saving network communication but is also utilised to manage queries and responses during periods of disconnected operation.

**Federation** The distributed brokers (device-based MCxBs and NCxBs deployed in the cloud infrastructure) collectively form an overlay network of brokers that manage local clients (consumers and providers). This federation of context brokers is achieved with a coordination model that is based on routing of subscriptions and notifications across distributed brokers, discovery and lookup functions and is described in detail in Section 3.3.

The MCxB offers these functions to device based context providers (MCxPs) and consumers (MCxCs) by exposing well defined interfaces. Externally, network based clients (NCxPs and NCxCs) can communicate directly with the mobile broker but device mobility causes changes in communication end points and hinders reachability of the MCxB (and hence its clients). This issue can be addressed by updating all external clients whenever the communication end point of the MCxB changes due to mobility. A better approach is to communicate with the external providers and consumers via a network-based broker (NCxB). Our prototype implementation uses the latter approach by federating the mobile and network brokers together into an overlay network of brokers. Clients of a broker only communicate with the local broker and queries, subscriptions and notifications are routed between brokers using a publish/subscribe communication paradigm. This model is described later in Section 3.3, a detailed theoretical model is also provided in our earlier work [Kiani et al., 2010a].

## 3.2 Data Model

The data model specifies the format of the communication and coordination that takes place between context consumers, providers and the brokers. While the information content is dependent on the domain and types of consumers and providers, a particular sub-set needs to be specified in order to utilise the brokering functions defined above. Clients of a broker are described using name-value attribute pairs and this description is used for registration with the broker. Subscription and notification further require that the acquired context be annotated with meta-information that allows categorisation and matching of context into specific instances that can be compared with subscriptions and queries. Korpipää et al. [2003] present some guidelines of designing information models that include properties of simplicity, flexibility, extensibility, generalizability and expressiveness. XML schema and ontology based information modelling approaches provide good coverage of these properties to varying degrees. In our architecture, an XML based schema for coordination and communication of context information, titled *ContextML*. ContextML specifies the model for context information, context subscription/notification and control messages as well. The following paragraphs describe the core elements of ContextML. A detailed description is provided in our earlier work [Knappmeyer et al., 2010].

**Entity** Every exchange of context data is associated with a specific entity, which can be a complex group of more than one entity. An entity is the subject of interest (e.g. user or group of users) which context data refers to and it is composed of two parts: a type and an identifier. The type refers to the category of entities e.g. username (for human users), IMEI (for mobile devices), SIP URI (for SIP accounts), room (for a sensed room) and group (for groups of other entities e.g. usernames or IMEI numbers). The entity identifier specifies a particular instance in a set of entities belonging to the same type. Every human user of the system can be related to many entities in addition to the obvious type username, therefore a component that provides identity resolution is necessary. A User Profile CxP is provided in the Context Provisioning Architecture to perform this function.

**Scope** Specific context information in ContextML is defined as scope and is a set of closely related context parameters. Every context parameter has a name and belongs to a certain scope. Using scope as context exchange unit is useful because parameters in that scope are always requested, updated, provided and stored at the same time; it means that data creation and update within a scope are always atomic and that context data in a scope are always consistent. Scopes can be atomic or aggregated in a union of different atomic context scopes. Examples of context scopes of varying degree of complexity include location, weather, activity, situation, cellular, and Wi-Fi network identification. Figure 4 illustrates the entity-scope association.

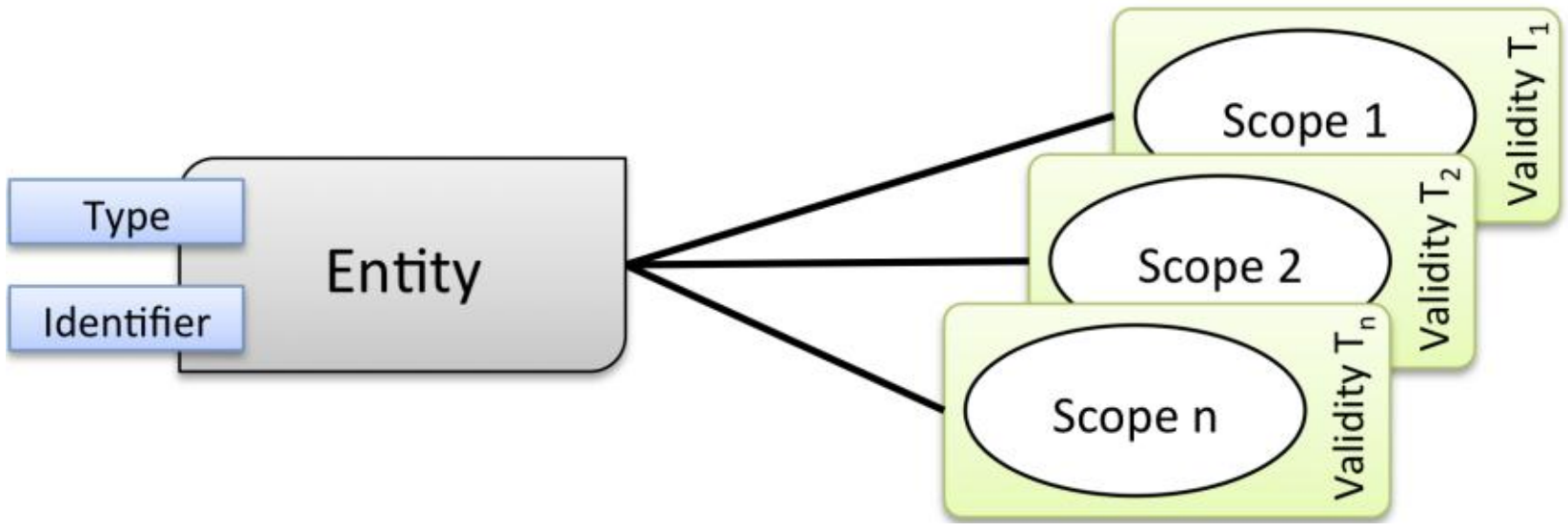

Figure 4: Entity–scope relationship, an entity can have many context scope associations. Scope instance have validity periods.

## 3.3 Context Data Representation

Whenever a CxC requests or subscribes to a specific context scope, it receives a response encoded in the ContextML schema element *ctxEl* when context is available. *ctxEl* contains information about where the context has been detected and encoded (CxP), which entity it is related to (*entity*), what *scope* it belongs to, and the actual context data in the dataPart element. A graphical description of this element, along with ContextML schema elements of context subscriptions (cxtSubscr) and notifications (cxtPublish) is given in Fig. 5. The elements par, parS and parA (Fig. 5) are simple constructs to store name-value pairs and attributed collections (arrays and structures) respectively. Every scope instance (context information) is tagged with a specific timestamp (time of context detection) and an expiry time. The expiry tag states the validity period of the scope after which the information is considered invalid.

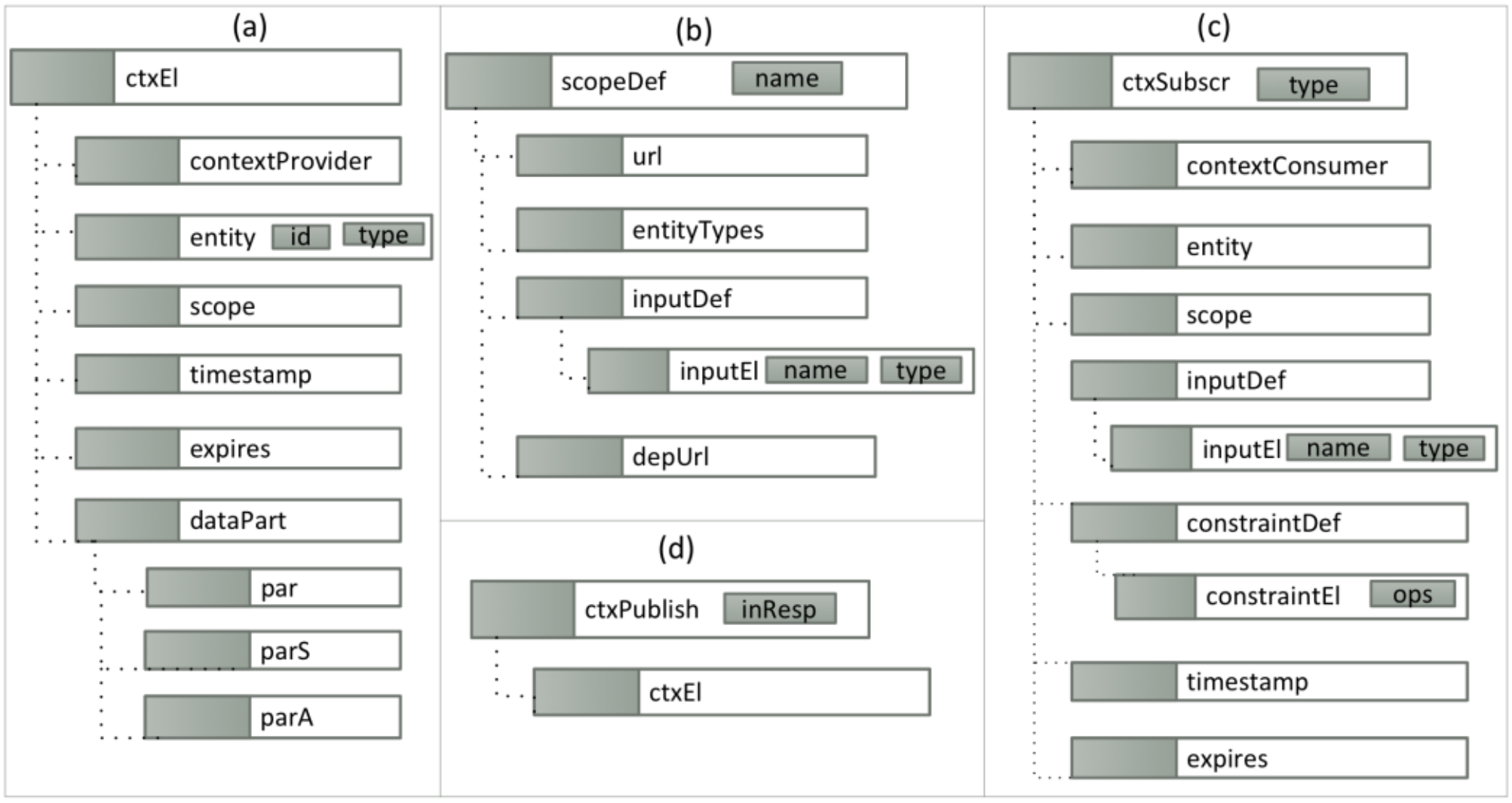

Figure 5: A subset of Context ML schema elements: a) Representation of context information in ContextML format (*ctxEl*). b) ContextML scope definition schema specifying the input parameters required for querying this scope, entity types for which this scope is valid and dependency on other scopes. c) ContextML subscription schema element (*ctxSubscr*) and d) ContextML notification schema containing *ctxEl*. For brevity, only essential attributes of the elements are shown.

**Asynchronous Event-based Query** This mechanism allows a context consumer to utilise an event-based publish/subscribe function in the broker for context queries and responses. In addition, a subscription can specify if the response is to be routed back via context brokers or direct call back to the subscribing consumer. Direct call back is suitable for non-mobile context consumers but makes the directly notified context data non-cacheable at the broker. Figure 5-c shows the ContextML schema of a context subscription. The context subscription by consumers is complemented by context notifications from the context providers. On availability of information that matches a certain subscription, a context provider generates a context notification (Fig. 5-d). The notification is routed to the subscriber CxC either directly or via context broker depending on the type of subscription. The subscriptions and notifications between clients of multiple brokers are managed through a coordination model that is described later in Section 3.3. The experimental evaluation discussed in Section 5 uses only this asynchronous form of the context query model.

**Synchronous Direct Invocation** Context consumers can formulate a simple context query for a particular context scope by invoking the context provider over HTTP and encoding the request parameters in the HTTP URL directly. The Lookup function in the broker is used by the consumers to find the communication endpoint of the relevant context provider, after which the query and response takes place between consumers and providers directly without any participation of the broker.

**Proxy Query and Subscription** A facility for context query and subscription provided by the broker is the proxy query mechanism. The proxy query service in the broker queries the required providers on behalf of the consumers when the requested context scope in the query or subscription is dependent on other scopes. Instead of requiring the context consumers to query each dependent scope, the broker satisfies the scope dependencies by querying for the dependent scopes when it receives a query about a context scope that is dependent on other scopes e.g. weather context of a user is dependent on the location scope. This mechanism is especially useful for resource constraint context consumers (e.g. context-aware applications executing on mobile phones) and becomes efficient if the broker is equipped with a local context cache that stores context data until its expiration.

## 3.4 Coordination Model

A real world deployment of a broker based context aware system may incorporate context providers and consumers that are geographically distributed. To reduce management and communication overheads, it is desirable to have multiple brokers in the system divided into administrative, network, geographic, contextual or load based domains. Context providers and consumers may be configured to interact only with their nearest, relevant or most convenient broker. But this setup demands inter-broker federation so that providers and consumers attached to different brokers can interact seamlessly. To achieve this, a simple event system can be implemented by an overlay network of distributed brokers for relaying subscriptions and notifications. Our system model for an overlay network of brokers working in a federation is based on the model presented by Mühl et al. [2006] and has been extended for context subscription/notification with the use of client advertisements. The conceptual development of this model is presented in the following sections.

### 3.4.1 System Model

The system model consists of a set of cooperating brokers that are arranged in a topology that is restricted to be acyclic. Each broker $B_i$ manages a mutually exclusive set of local clients $L_{B_i} = \{\kappa_1, \kappa_2, \ldots, \kappa_n\}$ and $L_{B_i} \subset K$ where $K$ is the set of all clients in the system. The clients here refer to CxCs and CxPs. Each broker $B_i$ is connected to a set of neighbouring brokers $N_{B_i} = \{\eta_{i_1}, \eta_{i_2}, \ldots \eta_{i_n}\}$ and $N_{B_i} \subseteq \mathcal{B}$ where $\mathcal{B}$ is the set of all brokers in the system.

### 3.4.2 Subscriptions

A subscription $\sigma$ contains a stateless logical expression that is applied to a notification $\nu$, i.e. $\sigma(\nu) \longrightarrow (true, false)$. A subscription can be given as a logical expression that consists of predicates that are combined by Boolean or logical operators (and, or, not, >, =, etc.). Such operators can be used to impose constraints while defining subscriptions (e.g. attribute name=”weatherCondition”). Consider an attributed subscription that imposes a constraint on the value of a single attribute, e.g. age > 25. The subscription constraint can be defined as:

$$\gamma_i = (n_i, op_i, C_i) \tag{1}$$

where $n_i$ is the attribute name, $op_i$ is a test operator and $C_i$ is a set of constants that may be empty. The name $n_i$ determines which attribute the constraint applies to. If a notification does not contain attribute named $n_i$ then $op_i$ evaluates to false. A notification *matches* $\sigma$ if $\sigma(\nu)$ evaluates to true. The set of matching notifications $N(\sigma)$ is defined as $\{\nu | \sigma(\nu) = true\}$.

### 3.4.3 Notifications

The broker exposes two interfaces namely $pub(Notification\ \nu)$ and $sub(Subscription\ \sigma)$ that allow the clients to publish or subscribe to events. The broker uses a $notify(Notification\ \nu)$ message itself to deliver notifications to local clients. Moreover, it uses a message $forward(Notification\ \nu)$ to forward notifications to neighbouring brokers (brokers who have clients subscribed for the current notification).

### 3.4.4 Client Registration Tables

Each broker $B_i$ maintains a client registration table $R_{B_i}$, which contains entries about its registered clients. A client $\kappa_i$ registers with a broker by providing an advertisement that contains a unique identifier $I_\kappa$ and information about its communication endpoint $U_\kappa$. In case the client is a CxP, the advertisement also contains the context *scope* served by the client. Neighbouring brokers exchange client registration tables amongst each other at regular intervals $\triangle X_R$. Out of turn triggering of client registration table update at a broker can occur when a new client registers with the broker so that availability of a new client is immediately disseminated in the system.

## 3.5 Subscription Tables

Each broker $B_i$ maintains a subscriptions table $T_{B_i}$, which contains entries about subscriptions related to its clients. Each entry in $T_{B_i}$ is a pair $(\sigma, D)$ consisting of a subscription $\sigma$ and a destination client $D \in \kappa \cup N_B$. Hence each broker maintains subscription entries only for its local clients and neighbouring brokers and not of the whole system entities. When a client $\kappa_j$ issues a subscription $\sigma_k$ to the broker $B_i$ that it is registered with, $B_i$ adds an entry $(\sigma_k, \kappa_j)$ to its subscriptions table $T_{B_i}$. Using the client registration table $R_{B_i}$, it determines which broker $B_s$ can satisfy the subscription and updates $B_s$ with the new subscription entry, which adds the entry $(\sigma_k, \kappa_j)$ to its subscription table $T_{B_i}$.

In the following sections, we describe how this model operates for one and two brokers in the system; the general case of $n$ number of brokers can be referenced from our earlier work Kiani et al. [2010a]. It is assumed in the following discussions that the clients have already registered with their respective brokers and, in case of two brokers, the brokers have already exchange client registration tables i.e. brokers *know* which client/broker can satisfy a subscription pertaining to a particular scope.

## 3.6 Single Broker Case

The local client $\kappa_1$ of broker $B_1$subscribes with the broker with subscription $\sigma_1$ using the $sub(Subscription\ \sigma_1)$ broker interface (Fig. 6). The broker saves this subscription in its subscription table and then determines that the local client $\kappa_2$ is capable of producing information that can satisfy the subscription. Broker $B_1$ forwards the subscription $\sigma_1$ to the local client $\kappa_2$ . The client $\kappa_2$ monitors its produced data in case it matches any of the subscriptions it has received via the broker. If and when subscription $\sigma_1$ is satisfied, $\kappa_2$ produces a notification $\nu_1$ and sends it to the broker via the $pub(Notification\ \nu_1)$ broker interface. The broker consults its subscription table $T_{B_1}$ and notifies the client that has the relevant subscription entry, in this case $\kappa_1$.

## 3.7 A Case of Two Brokers

Consider this case with the help of Fig. 7 where the local client $\kappa_1$ of broker $B_1$ subscribes with subscription $\sigma_1$ . Broker $B_1$ saves the entry $(\sigma_1, \kappa_1)$ in its routing table $T_{B_1}$, which was initially empty. It then sends the following message (table exchange) to its neighbouring broker $B_2$:

$$subTableUpdate\,(B_1, \sigma_1) \tag{2}$$

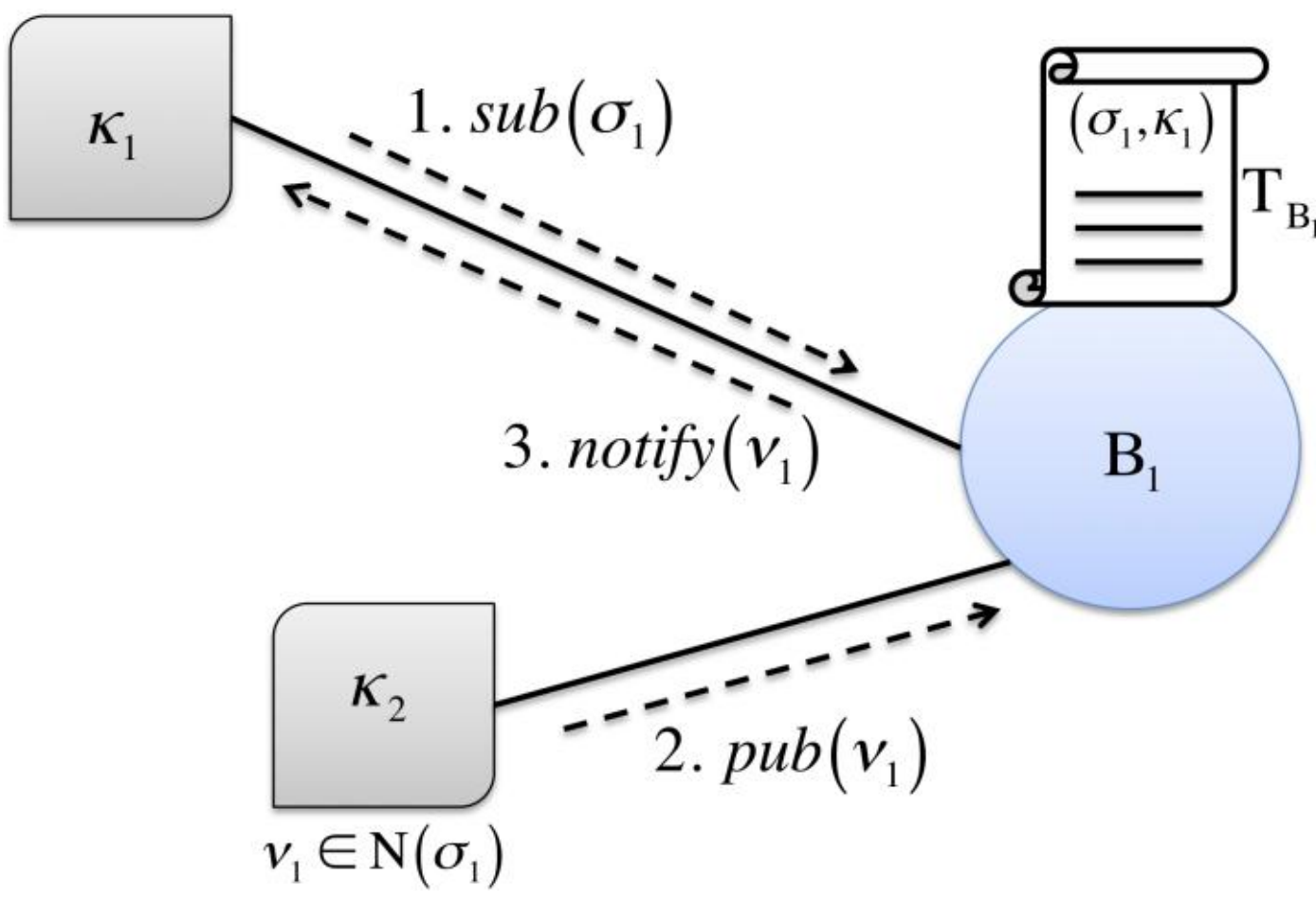

Figure 6: Single broker coordination scenario.

This causes the broker $B_2$ to update its routing table with the entry $(\sigma_1, B_1)$. Broker $B_2$ has two registered clients $\kappa_2$ and $\kappa_3$. Broker $B_2$ forwards $\sigma_1$ to $\kappa_2$ considering it to be a source of matching notifications for this subscription by consulting its client registration table $R_{B_2}$ and evaluating the *scope* entries in client advertisements, e.g. if the subscription is regarding weather updates of a certain area, then only a CxP that produces weather related context may be forwarded the subscription information; it may not be relevant to forward a weather related subscription to a client that produces context about proximity of a group of users. When $\kappa_2$ produces information that satisfies $\sigma_1$, it sends a notification $\nu_1$ to $B_2$ along with the information that this notification satisfies the subscription $\sigma_1$, i.e. $\sigma_1(\nu_1) \longrightarrow true$. The broker $B_2$ analyses its subscription table $T_{B_2}$ and finds entry $(\sigma_1, B_1)$ and therefore forwards the notification $\nu_1$ to $B_1$:

$$forward(B_1, \nu_1) \quad (3)$$

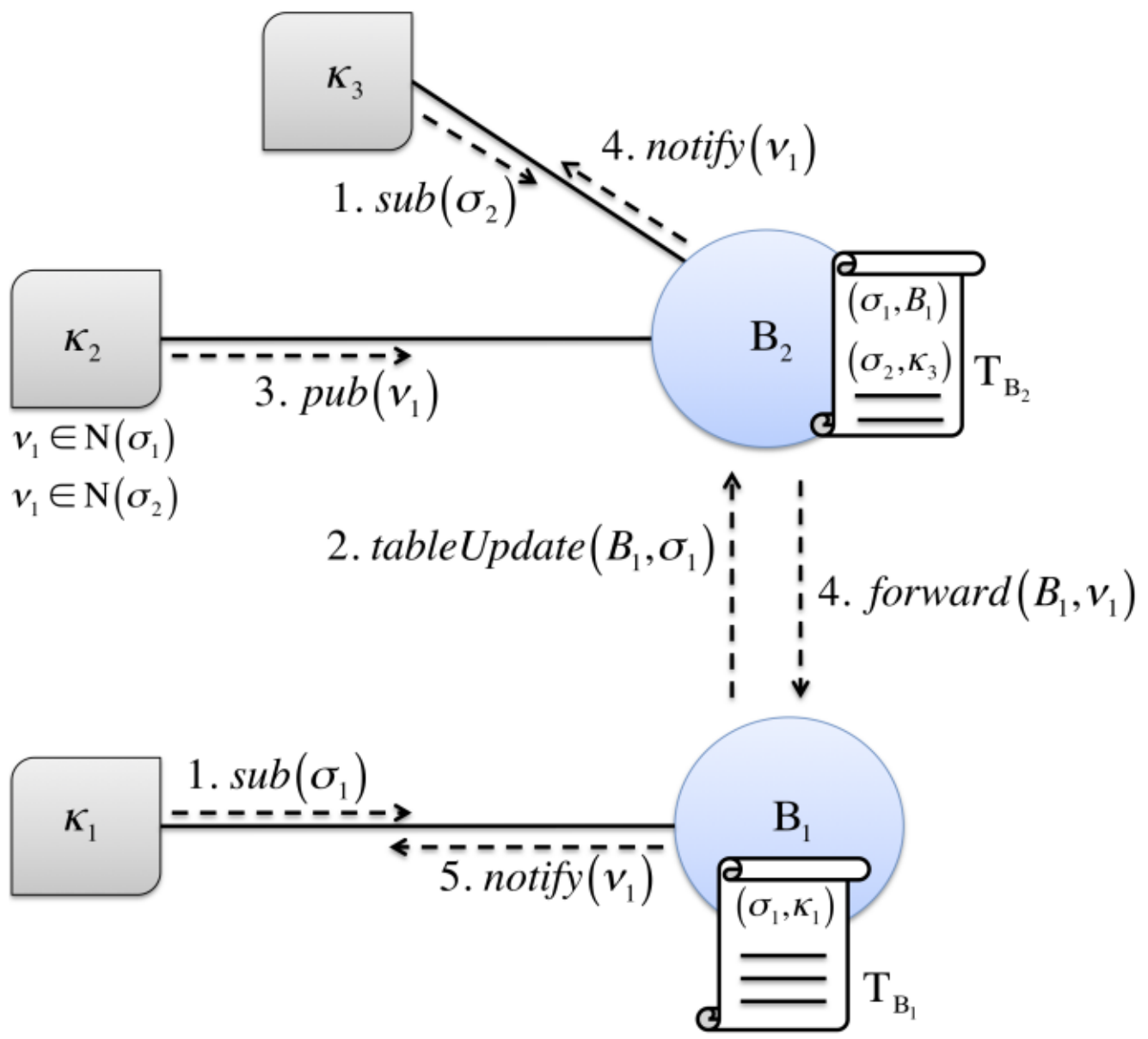

Figure 7: Two-broker coordination scenario.

The client $\kappa_1$ is a local client of the broker $B_1$ . Therefore $B_1$ uses the $notify(\nu_1)$ procedure to notify the client with the notification $\nu_1$. Consider an additional subscription $\sigma_2$ received from the local client $\kappa_3$ of broker $B_2$, which is also satisfied by $\kappa_2$. In this case, the subscription routing table $T_{B_2}$ will contain an additional entry $(\sigma_2, \kappa_3)$. Assuming that the notification $\nu_1$ produced by $\kappa_2$ evaluates to true for both $\sigma_1$ and $\sigma_2$, $B_2$ will calculate the set of matching destinations as:

$$\Gamma_{B_2}(\nu_1) = \{B_1, \kappa_3\} \quad (4)$$

for the notification $\nu_1$. For the local client $\kappa_3$, $B_2$ will invoke $notify(\kappa_3, \nu_1)$ locally. The other match $B_1$ is a remote broker and $B_2$ will invoke Eq. 3. For the local client $\kappa_1$ of the broker $B_1$, $B_1$ will then invoke $notify(\kappa_1, \nu_1)$ locally.

This section has described the coordination model via which brokers facilitate their local clients in exchanging context subscriptions and notifications across the overlay network of context brokers. The brokers discussed in this coordination model can be network level brokers, deployed on hosts in the network, or mobile brokers operating on user devices. In the following section, we describe the design elements in the Context Provisioning Architecture that effect energy consumption on mobile devices. Thereafter, we describe an experiment and associated results that signify the benefits of the Mobile Context Broker in energy conservation on mobile devices with a number of context consumers and providers.

# 4 Design Elements Effecting Energy Consumption

The design of the Context Provisioning Architecture, specifically the MCxB component and associated coordination model, leverages several benefits in the area of energy conservation. These benefits are possible due to the following aspects of the design:

**Non-blocking context query and response** Due to the asynchronous coordination model, consumers and providers on the mobile device are not blocked while waiting for their subscriptions and notifications to reach other providers and consumers (whether in the cloud or on the device). These components can continue performing other tasks or wait without being blocked (idle state), which results in lesser overall execution load on the device, thus conserving energy.

**Lightweight Consumers and Providers** Consumers and providers are only required to communicate with one local component (mobile broker) and are not concerned with communicating to a variable number of local and remote components. Their functional tasks are therefore limited and simplified by the coordination and communication facilities offered by the MCxB. Reducing this functional complexity of components results in lightweight context consumers and providers on the mobile device that use lesser computation power and thus conserve energy.

**Avoidance of Repetitive Tasks** Various tasks that are otherwise repetitive e.g. subscription routing, notification forwarding and, network connectivity monitoring, are delegated to the MCxB in this design and this aspect serves as another factor in conserving energy.

**Bulk-mode Communication** The bulk query mode of the MCxB reduces over-all network communication and potentially conserves energy. This mode is specifically evaluated in one of the experiments and the results are described in Section 5.

**Local Cache** The MCxB maintains a cache of recently produced and received context notifications. Depending on the cache-hit rate, the caching facility reduces not only the overall context related traffic but also saves computation cycles by not invoking a context provider to satisfy a subscription.

In the reference implementation of the Context Provisioning Architecture all cloud based context brokers provide HTTP communication interfaces to their clients. The provisioning of these communication facilities is mirrored in the MCxB implementation. But there is a significant difference between the manner in which clients of a MCxB interact with it as compared to the case of network context brokers i.e. clients of the MCxB execute on the same computing device as their local broker in contrast to the network based clients of the context brokers deployed in the cloud infrastructure, which are more likely to be executing on computing systems separate from that of the broker e.g. desktops. This situation provides the opportunity to exploit native Inter-Process Communication (IPC) facility that are available on our reference mobile device platform i.e. Android. But such a provision raises the possibility of incompatibility between device based consumers and providers using native IPC and rest of the system that utilises a standardised distributed communication protocol i.e. HTTP. This issue is inherently resolved in our design as the device platform native communication mechanism is only used between the device based clients and the Mobile Context Broker and any external communication always takes place over standardised communication protocols (HTTP). Our primary reason for providing native IPC based communication facility for device-based components is *efficiency*. Efficient exchange of messages or processing of inter-process calls can be achieved by using a communication mechanism that is lightweight and uses least resources (e.g. creation of threads to process individual requests, allocation of buffers in the memory). Our analysis has revealed that on our reference implementation platform (Android SDK), native IPC calls between two processes take an order of magnitude less time to complete than HTTP and TCP/IP socket based calls. Figure 8 illustrates the results of our analysis during which a number of calls were made between two processes on an Android based device using different communication mechanisms (HTTP, TCP sockets and IPC). A fixed ContextML encoded request and response message was exchanged between the two processes and the completion time of each call was recorded.

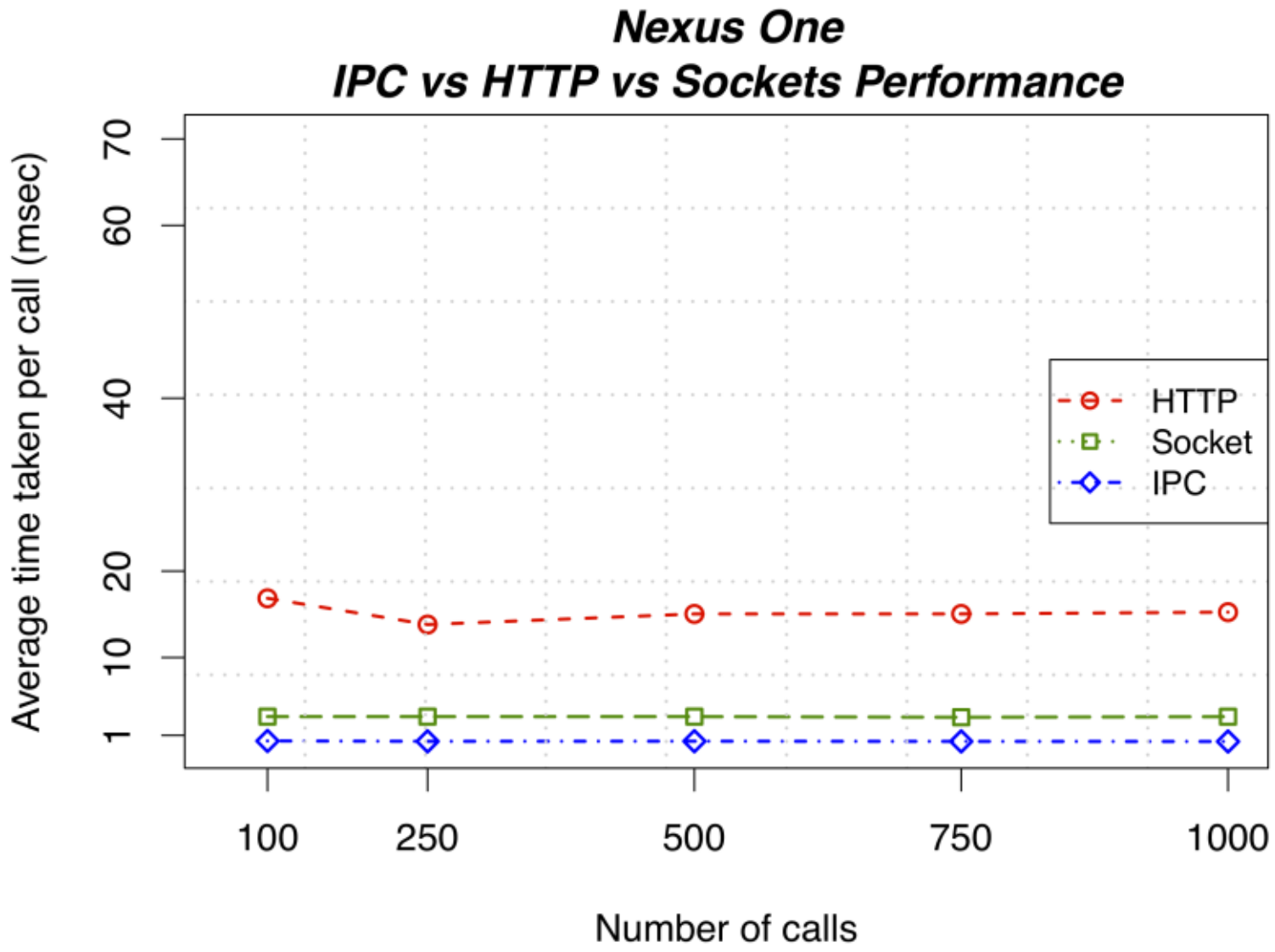

Figure 8: Mean completion time of calls between two processes executing on an Android based Google Nexus One device using different communication technologies.

The results show that IPC calls complete at least three times faster than calls over TCP/IP sockets and at least 15 times faster than HTTP calls. We have also performed comparative analysis on devices with different hardware configurations and concluded that the differences observed in case of the first device are also observed in other devices e.g. Fig. 9 shows the results of the same experiment on an HTC Wildfire phone, which has a less capable CPU than the device used in the earlier experiment. The difference in performance of IPC, HTTP and socket based communication mainly arise due to their different semantics and amount of implicit I/O operations i.e. I/O operations not requested explicitly by the process using the facility e.g. creation and allocation of memory buffers and creation of separate threads to process an invocation or request. The semantics of IPC, TCP/IP sockets and HTTP are largely standardised and do not differ by much between different versions of the Android SDK e.g. results of the same experiment on a Google Nexus phone based on Android SDK version 2.1 only differ by a maximum $\pm 2\%$ from those of the same phone based on Android SDK version 2.3.

The results of this experiment show that there is a notable difference in the time taken by different communication mechanisms to exchange the same piece of data. The improvement offered by the native IPC mechanism is significant enough to warrant the provision a separate local communication mechanism for components executing on the mobile device. Communication with cloud-based brokers continues to take place over HTTP interfaces. This observation i.e. efficiency of IPC based communication over HTTP in our setup, is exploited in our architecture and the implications will be discussed further in Section 6.4.

The practical impacts of the design elements and the evaluation carried out to analyse the effect of the Mobile Context Broker on energy consumption on a mobile device are discussed in the following section.

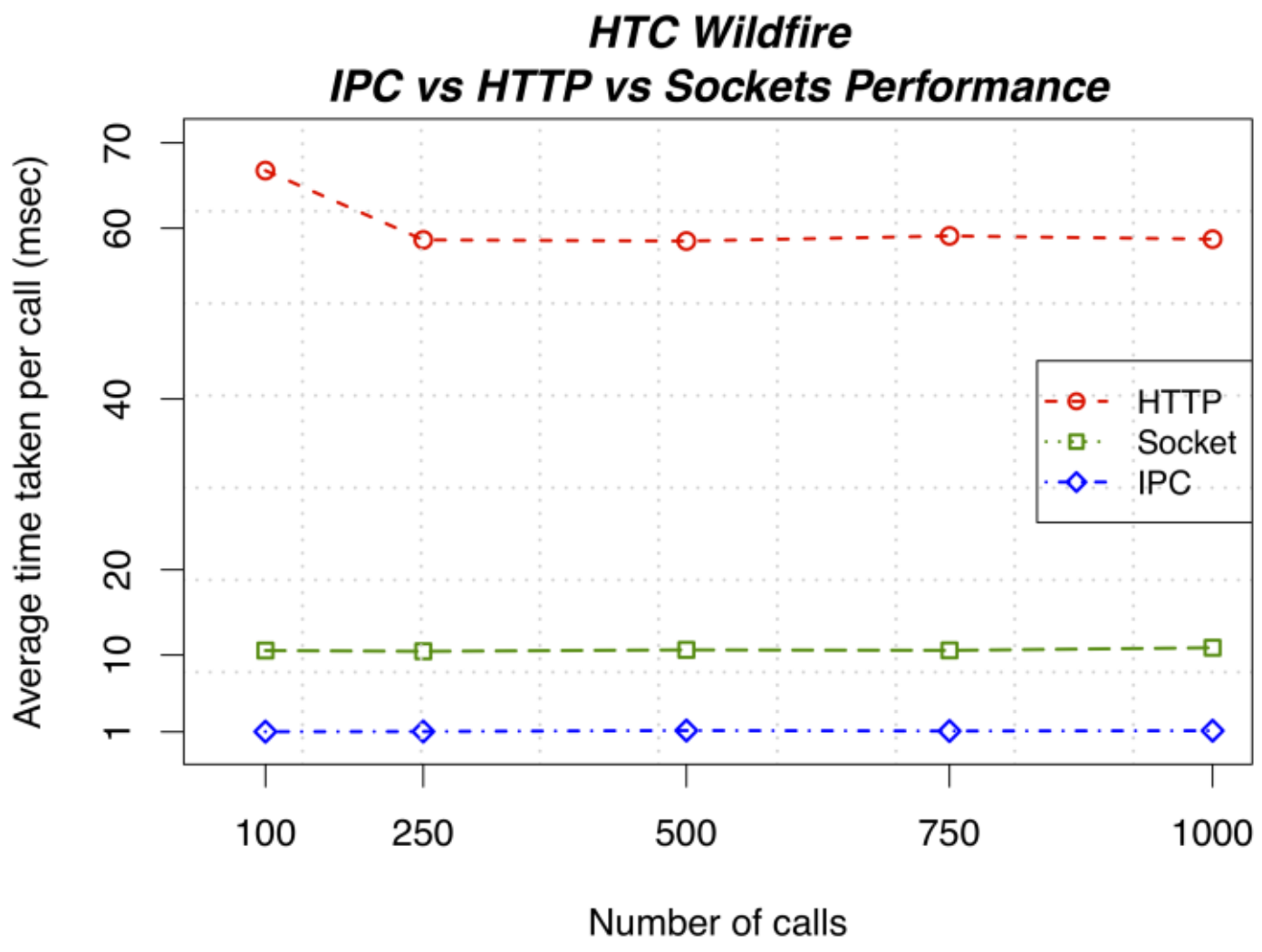

Figure 9: Mean completion time of calls between two processes executing on an Android based HTC Wildfire device using different communication technologies.

# 5 Evaluating Energy Consumption with the Context Broker

We have used the concept of the Context Broker presented in the preceding section for coordination and dissemination of user and environmental contextual information. Context information acquired at the mobile device by context providers or requested by the mobile device based context consumers from external context providers is brokered via this component. The following paragraphs discuss the prototype MCxB from an implementation point of view and present the scenario used to determine the benefits of employing such a broker for context coordination and communication between mobile devices and cloud services.

## 5.1 Mobile Context Broker Prototype

The prototype Context Broker is developed on the Android platform (www.android.com) and executes as a background service. Various mobile device based context-consuming applications that have been developed include context-based content providers (news, entertainment etc.), location based gameplay and context-based shopping recommendation applications. Mobile device based context providers include location, proximity activity providers. Various context providers are deployed in the cloud and include weather provider, user profile and preferences provider etc. In addition the mobile device acts as a gateway for nearby environment and wearable sensors for dissemination of their sensed information). Development is underway to use Bluetooth enabled, wearable air quality sensors for participatory sensing and using the mobile broker to disseminate collected air quality information.

On the network side, the context broker NCxB is based on JavaEE technology and exposes interfaces to context consumers and providers in the cloud. All communication between the brokers and their clients takes place over HTTP with ContextML encoded messages, subscriptions and notifications. The choice of these standardised technologies provides interoperability and allows for interaction between brokers and clients executing on heterogeneous hardware. A significant feature of the broker is the caching facility. No matter what type of context information, it remains valid for a certain amount of time e.g. weather, user activity, and user preferences. We utilise this temporal property of context information by annotating a validity period in context meta-data. A broker uses this validity period to cache recently retrieved information. We have reported an empirical study on advantages of caching facility in a context broker in our earlier work Kiani et al. [2010b], which establishes significant improvements in query response time and reduction of network bound traffic.

## 5.2 Experiment Scenario

In order to experimentally analyse the benefits of using a context broker on a smart mobile device, a scenario is designed to monitor various parameters with and without the use of the MCxB. Ten context consumers and five context providers are deployed on the mobile device. Similarly, five context consumers and five context providers are deployed in the network simulating the cloud infrastructure. For simplicity each context provider provides a unique type of context and each context consumer is only interested in one type of context available at one of the context providers. All queries and responses are in the form of subscriptions and notifications. Table 1 lists the parameters used in the setup.

Table 1: Experiment setup parameters

| Parameter | Value | | | |
|---|---|---|---|---|
| | **Mobile Device** | | | |
| Number of Consumers | 10 | | | |
| | *Consumer ID* | *Scope of Interest* | | |
| | MCxC_1 | devScope_1 | | |
| | MCxC_2 | devScope_2 | | |
| | MCxC_3 | devScope_3 | | |
| | MCxC_4 | devScope_4 | | |
| | MCxC_5 | devScope_5 | | |
| | MCxC_6 | networkScope_1 | | |
| | MCxC_7 | networkScope_2 | | |
| | MCxC_8 | networkScope_3 | | |
| | MCxC_9 | networkScope_4 | | |
| | MCxC_10 | networkScope_5 | | |
| Number of Providers | 5 | | | |
| | *Provider ID* | *Scope ID* | *Response Size (bytes)* | *Validity (sec)* |
| | MCxP_1 | devScope_1 | 750 | 30 |
| | MCxP_2 | devScope_2 | 1000 | 60 |
| | MCxP_3 | devScope_3 | 1500 | 200 |
| | MCxP_4 | devScope_4 | 2000 | 350 |
| | MCxP_5 | devScope_5 | 5000 | 900 |
| Number of Brokers | 1 | | | |
| | **Network/Cloud Infrastructure** | | | |
| Number of Consumers | **5** | | | |
| | ***Consumer ID*** | *Scope of Interest* | | |
| | **MCxC_1** | devScope_1 | | |
| | **MCxC_2** | devScope_2 | | |
| | **MCxC_3** | devScope_3 | | |
| | **MCxC_4** | devScope_4 | | |
| | **MCxC_5** | devScope_5 | | |
| Number of Providers | **5** | | | |
| | ***Provider ID*** | *Scope ID* | *Response Size (bytes)* | *Validity (sec)* |
| | **MCxP_1** | networkScope_1 | 750 | 30 |
| | **MCxP_2** | networkScope_2 | 1000 | 60 |
| | **MCxP_3** | networkScope_3 | 1500 | 200 |
| | **MCxP_4** | networkScope_4 | 2000 | 350 |
| | **MCxP_5** | networkScope_5 | 5000 | 900 |
| Number of Brokers | 1 | | | |

Two experiments are carried out, one with the device-based broker (MCxB) and the second without it. On the cloud side a context broker NCxB is deployed with five context consumers and five context providers with scopes shown in Table 1. When MCxB is available on the device, device context consumers and providers register with the local mobile broker. MCxB itself discovers and exchanges registration information with the cloud broker NCxB. All subscriptions and notifications are exchanged via the context brokers and not directly between consumers and providers. In the experiment scenario where the MCxB is absent, device based consumers and providers register with and use the remote network context broker NCxB. Each context consumer sends subscriptions concerning a scope of at most one context provider, i.e. context queries that are dependent on more than one scope are not used in the experiment. The experiment deployment structure is shown in Fig. 10.

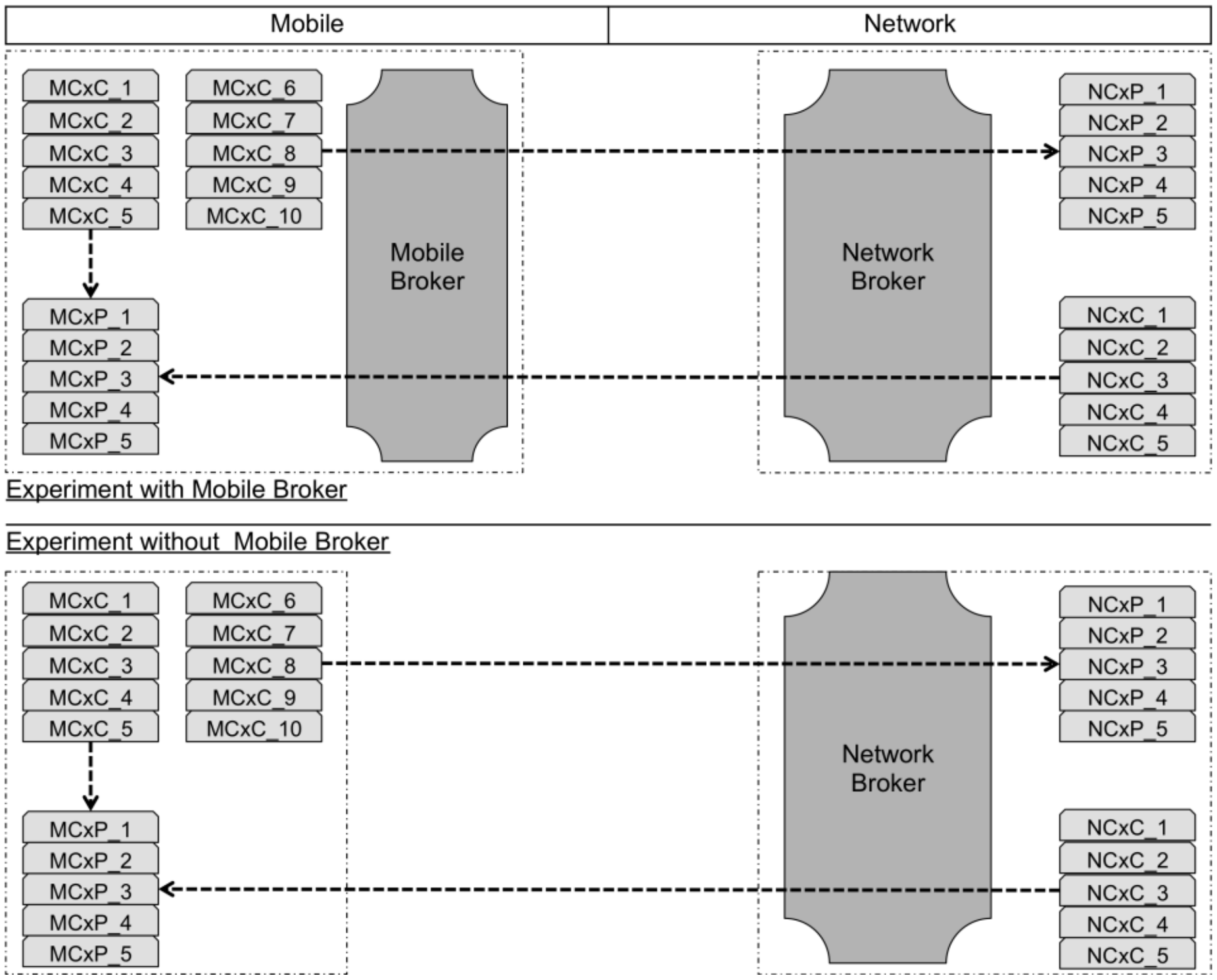

Figure 10: Deployment structure of the experiments. Arrows show the interest of CxCs in particular CxPs.

The rate at which the consumers query (send subscriptions) is determined by an exponential distribution. The process of the arrival of new query can be seen as a Poisson process in which events (queries) occur continuously and independently of each other with an average inter-arrival time of $\lambda$. This process is analogous to page view requests to websites, a well-modelled Poisson process (cf. [van Gelder et al., 2002]). The exponential distribution defines the time between events (queries/subscriptions in this case) in the Poisson process. In these experiments, $\lambda$=50 is chosen based on earlier findings [Kiani et al., 2010b] that showed that faster rate of requests than this does not affect the query satisfaction performance of the system.

In this experiment set, all subscriptions are one-time subscriptions, i.e. there is no need to send consecutive notifications. Providers respond to subscriptions with emulated context after a period of $d = UniformRandom(10, 2000)$ milliseconds to emulate delayed responses/processing time. This range is based on observed numbers form real world deployment of context providers and takes into consideration database/storage access times, processing, reasoning and encoding contextual data into a ContextML notification. The spread of various scopes in all queries is selected to be nearly uniformly distributed i.e. the experiments cater for the scenarios where the scopes are evenly distributed in all queries. The effect of context scope distributions that are biased towards a subset of scopes on query satisfaction times is discussed in [Kiani et al., 2010b].

Each experiment is repeated on the same set of hardware (mobile device and cloud server) and Wi-Fi interface is used on the server and mobile device within a single WLAN for minimising the effects of unpredictable network round-trip times. The mobile device used in the experiments is an Android version 2.2.1 based Nexus One smart phone (www.google.com/phone/detail/nexus-one). PowerTutor [Zhang et al., 2010] is used for calculating the energy used by individual applications (mobile broker, consumers and providers) on the device. PowerTutor is an application for Google phones that displays the power consumed by major system components such as CPU, network interface, display, etc. and different applications. This application allows software developers to see the impact of design changes on power efficiency. PowerTutor calculates the phone's breakdown of power usage with an average of 1% error over 10-second intervals while the worst case error over 10 seconds is 2.5% (cf. [Zhang et al., 2010, p. 8]). In these experiments only the energy used by an application in utilising the CPU and Wi-Fi is considered when calculating its energy consumption signature. All results in the following section are mean values of five repetitions of each experiment. Comparison of results from individual iterations shows variations within $\pm 3\%$.

# 6 Results

The purpose of the experiment is to find out the effects of using a broker for coordination and dissemination of contextual information collected at or required by the mobile device and cloud services. The following paragraphs discuss the results of the experiments described in the previous section.

## 6.1 Effect on Energy Consumption

Figure 11 shows the comparison of energy consumption on the mobile device with and without utilising a mobile broker. The experiment is repeated with 100, 1000, 2000 and 5000 queries exchanged between the device and cloud services. All queries are satisfied with a valid response form the appropriate context provider. In this case network availability is 100%, i.e. no disconnection takes place during the experiment. The results show that the use of the MCxB only results in energy conservation after a certain threshold of number of queries. Initially, for a small number of queries, the extra broker process results in increase in energy usage at the device. With the increase in number of queries over time, the caching feature of the MCxB saves significant energy by satisfying a portion of queries locally from cache instead of initiating network communication. The cache hit rate in these experiments varied from 17-20%. This hit rate is marginally less than that of around 35% reported in our earlier work [Kiani et al., 2010b]. The reason for this is that the experiments in that study are based on a network broker with greater resources, a larger number of providers and consumers registered with it and a larger number of context related entities (usernames, IMEI numbers etc.) present in the system

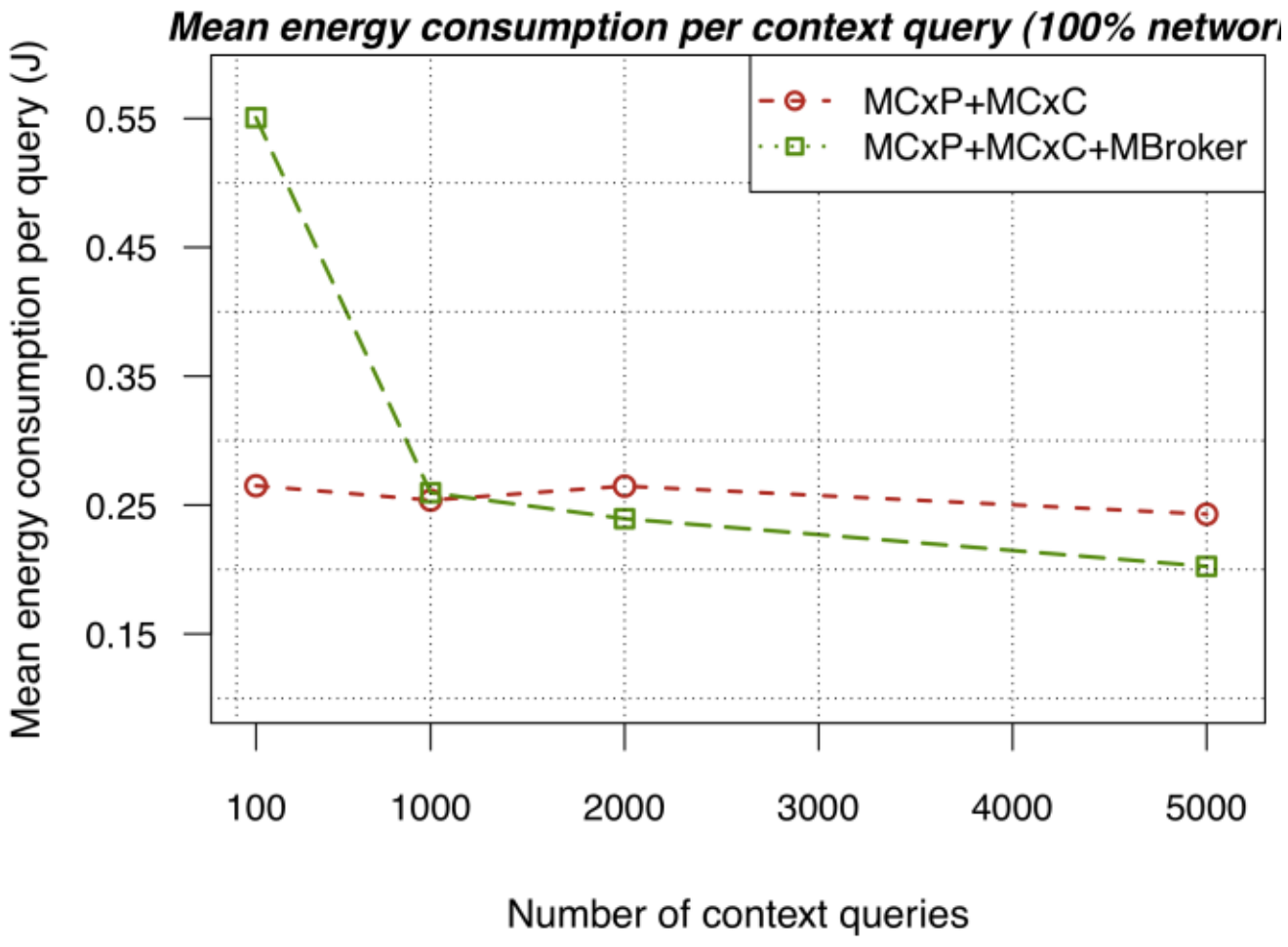

Figure 11: Energy consumption at the device with increasing queries and 100% network availability.

## 6.2 Varying Network Availability – Effect on Energy Consumption

The benefits of utilising a broker become pronounced during scenarios where network connectivity is intermittent. The chart in Fig. 12 shows energy conservation in scenarios where network availability varies form 100% to 50%. This experiment is carried out with 1000 queries each from cloud and device based context consumers. This figure is selected because energy consumption with and without the MCxB with 1000 queries and full network availability is almost similar (cf. Fig. 11). Because consumers and providers registered with the MCxB do not need to monitor network connectivity during periods of network unavailability, a portion of their execution cost is saved. The MCxB, which is responsible for routing queries and responses to and from the network for its local clients, manages the network communication and, combined with the local caching facility, it provides significant energy conservation. The mean energy consumption per context query reduces with decreasing network availability in case of the MCxB based setup.

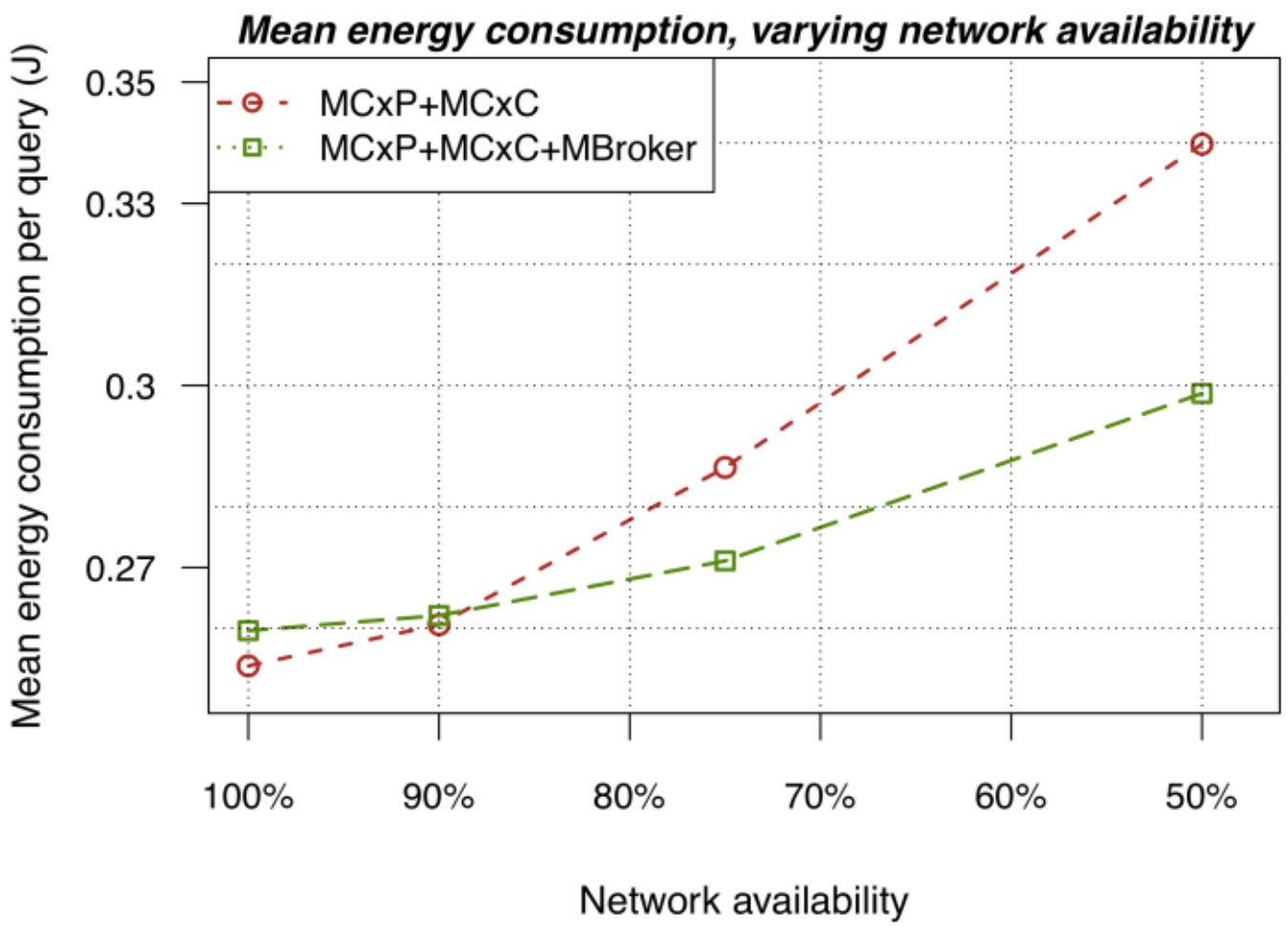

Figure 12: Energy consumption at the device during different levels of network availability.

## 6.3 Bulk Mode MCxB – Effect on Energy Consumption

During this experiment, the Mobile Context Broker is operated in bulk query mode i.e. the MCxB forwards subscription queries to the network in bulks of five queries. To operate in this mode, the MCxB examines the optional priority field in each subscription and if priority is set to low then the query is added to the bulk queue, which is only processed upon reaching a bulk limit (five in this experiment). A bulk queue is maintained for a maximum duration γ where γ is one half of the time remaining in expiration of the subscription with the earliest expiry time. The half limit is chosen so as to leave adequate time for response to reach the subscribing consumer. The duration γ is re-evaluated on addition of each low priority subscription to the bulk queue. A bulk queue is immediately processed either when γ is reached or number of queries reaches the pre-defined bulk limit.

Fig. 13 shows the energy consumed at the device under 100% network availability in comparison to the earlier experiment. While the MCxB operating in bulk mode still consumes less energy overall than non-broker scenario, the energy consumption increases in comparison to the case where the MCxB processes each subscription immediately. Any advantages of bulk query forwarding are offset by the increase in the duration for which the system has to operate in order to fully process all the queries. In addition, when queries are being processed in bulk, the mobile broker's cache takes longer to populate and cache hit ratio drops consequently (to 11.5% on average). A drop in cache-hit ratio signifies increase in network bound traffic as compared to the scenario where the MCxB processes subscriptions immediately. Moreover, this experiment is performed with all subscriptions set to low priority and hence the results shown in Fig. 13 signify the *maximum* benefit that can be garnered from bulk queries and lower percentage of bulk queries did not yield any more gains in energy conservation at the device.

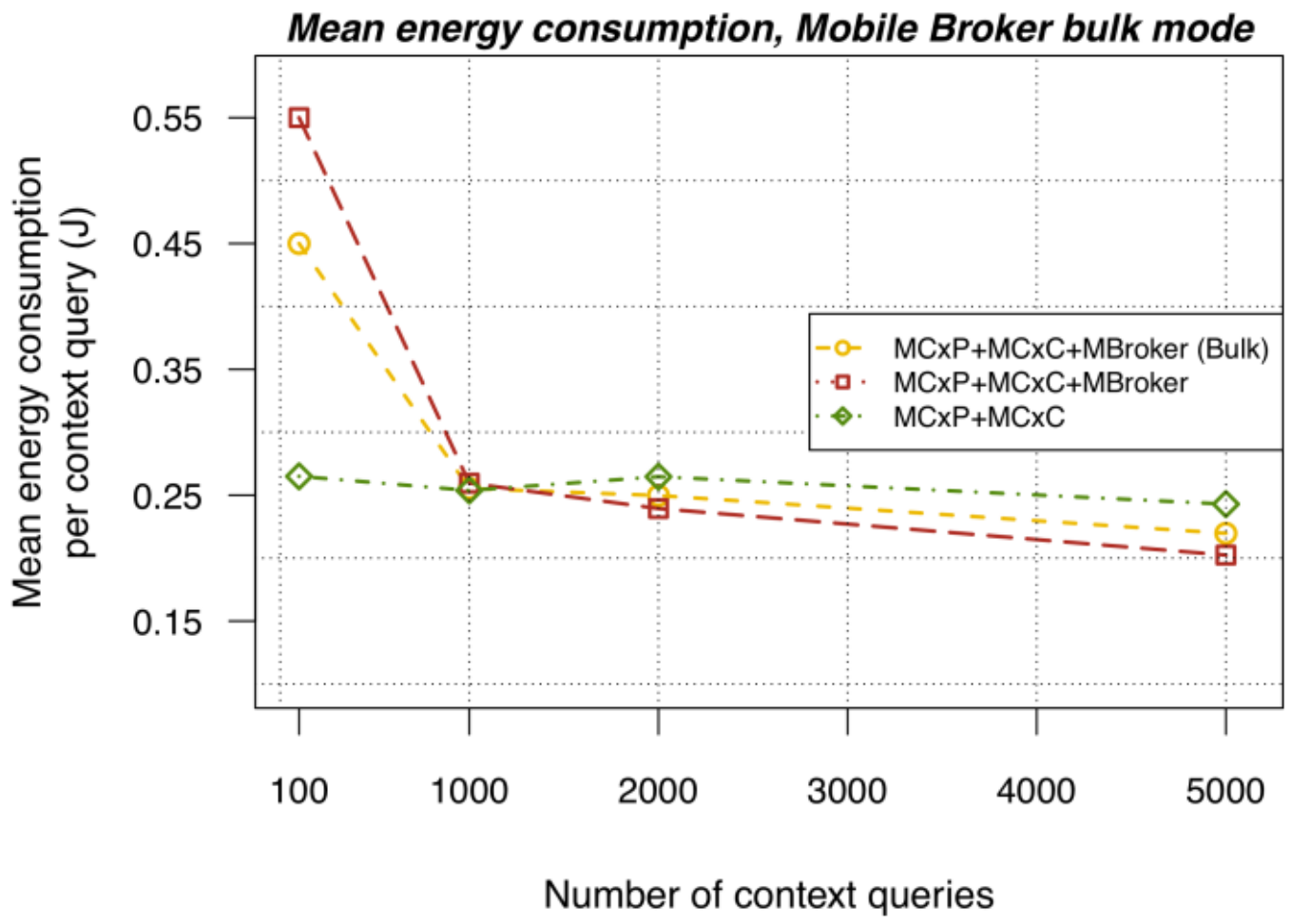

Figure 13: Energy consumption at the device with broker operating in bulk query mode.

## 6.4 Effect of IPC on Energy Conservation

Based on the observation in which we noted the mean time taken to complete a call using IPC is less than that of HTTP (cp. Fig. 8, Fig. 9), we hypothese that mean energy consumed per call also varies across different communication mechanisms. In order to verify this hypothesis, we record the energy consumed by different communication mechanisms while exchanging ContextML encoded messages between two processes executing on the same device. The results of this experiment, which is carried out on a Google Nexus One phone, are recorded in Table 2. The ‘client’ and ‘server’ data columns in Table 2 show the energy consumed by the components making the call and responding to it, respectively, while their sum is shown in the encompassing data column as well. The mean energy consumption measurements during 500, 1000, 2000 and 3000 request-reply calls using different communication mechanisms are plotted in Fig. 14.

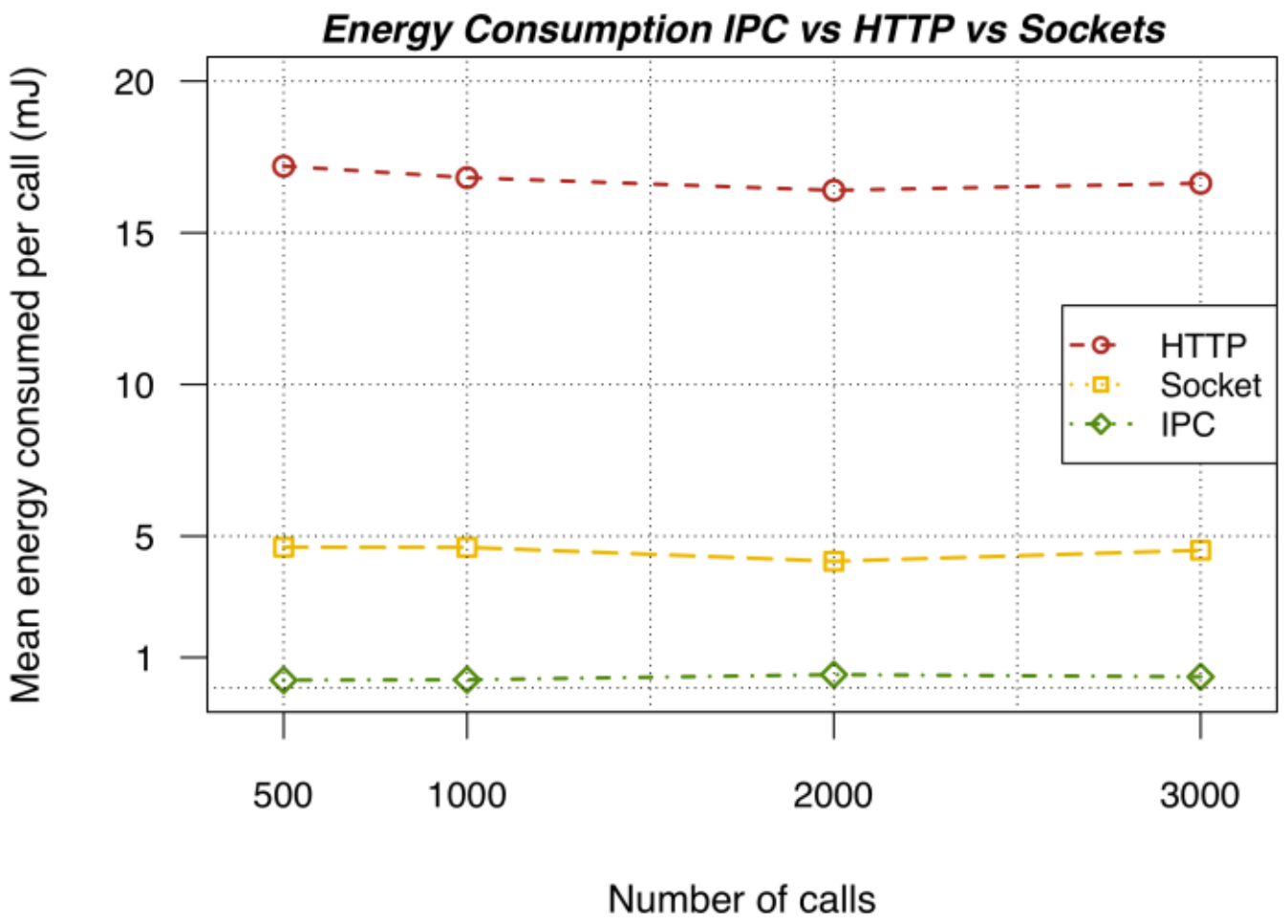

Figure 14: Mean energy consumption/call on Android Nexus One device – IPC vs. HTTP vs. Sockets.

The results show that IPC is significantly less energy consuming that HTTP and sockets based communication mechanisms, specifically on average an IPC call consumes almost 12 times less energy than a Socket based call and nearly 46 times less energy than an HTTP based call. One of the main factors for this divide is that IPC facilities are provided natively on the Android platform while creation of sockets, HTTP servers and related thread management has to be done by the developer (as carried out in our experimental setup), resulting in additional processing steps, creation of memory buffers etc.

Table 2: Total energy consumption on an Android Nexus One device - IPC vs. HTTP vs. Sockets calls.

| Number of calls | Total energy – IPC (J) | | Total energy – HTTP (J) | | Total energy – Sockets (J) | |
|---|---|---|---|---|---|---|
| | Server | Client | Server | Client | Server | Client |
| 500 | 0.092 | 0.034 | 4.08 | 4.52 | 1.32 | 0.998 |
| | 0.126 | | 8.6 | | 2.318 | |
| 1000 | 0.147 | 0.116 | 8.03 | 8.79 | 2.87 | 1.76 |
| | 0.263 | | 16.82 | | 4.63 | |
| 2000 | 0.554 | 0.311 | 15.65 | 17.15 | 4.82 | 3.52 |
| | 0.865 | | 32.8 | | 8.34 | |
| 3000 | 0.603 | 0.482 | 27.33 | 22.7 | 7.456 | 6.145 |
| | 1.085 | | 49.9 | | 13.601 | |

A notable observation in Table 2 is that different ratio of energy consumed by the server and client processes in the request-reply calls (plotted in Fig. 15a, 15b and 15c). In case of IPC and sockets, a slight majority (~60% and ~57% respectively) of the energy consumption is due to the server process while in case of HTTP it is almost equally distributed. The communication semantics in HTTP communication are analogous to those of sockets based communication i.e. the server process spawns a new thread to handle communication with a connecting client. But the communicated data in socket based communication, in our experiments and in general, is usually a byte stream of primitive and basic data types (int, double, char, String etc.), whereas in case of HTTP communication any data exchanged between client and server is further packaged into HTTP protocol specific packets. This packing and unpacking of data is an overhead that does provide the benefit of standardisation and interoperability, but renders HTTP communication more computationally and I/O intensive than sockets. Both client and server processes have to carry out an identical set of operations for communicating over HTTP, with spawning of a new thread to handle incoming request the extra step carried out by the server process. Due to these factors the server and client processes in HTTP communication consume an identical share of total energy but the overall energy consumption is much greater (see Fig. 14) than sockets and IPC based communication in these experiments.

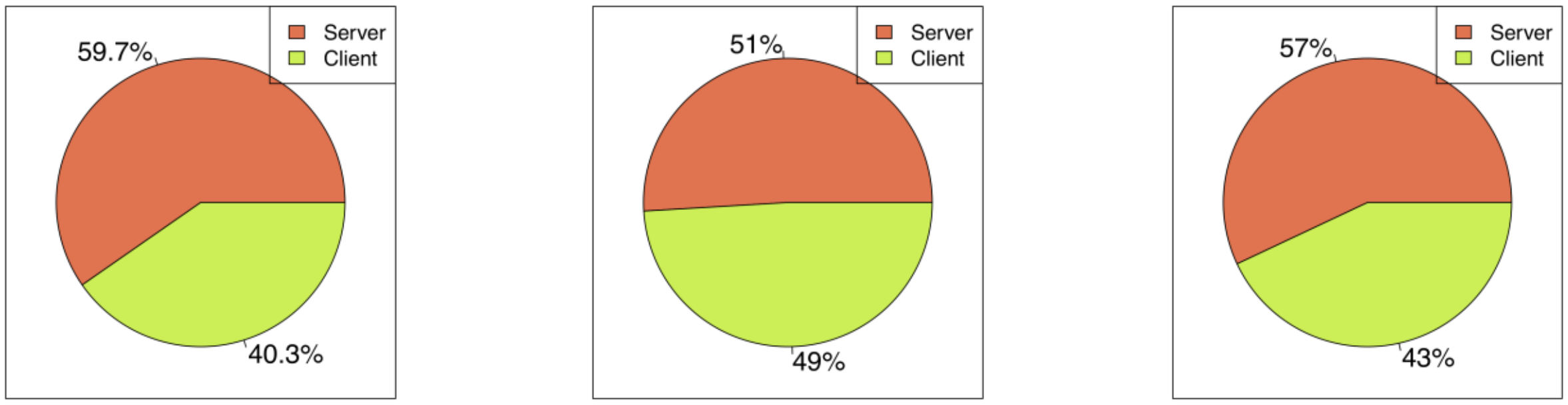

(a) IPC calls client-server energy consumption ratio

(b) HTTP calls client-server energy consumption ratio

(c) Socket calls client-server energy consumption ratio

Figure 15: Ratio of energy consumed by client and server process during IPC/HTTP/Socket calls.

## 6.5 Extended Results

The preceding discussion and results reinforce the usage of IPC based communication between consumers, providers and the prototype Mobile Context Broker in the Context Provisioning Architecture. Figures 16 and 17 show the extended results obtained by repeating the experiment scenario described in Section 5.2 and using IPC for device based consumers–mobile broker–providers interaction.

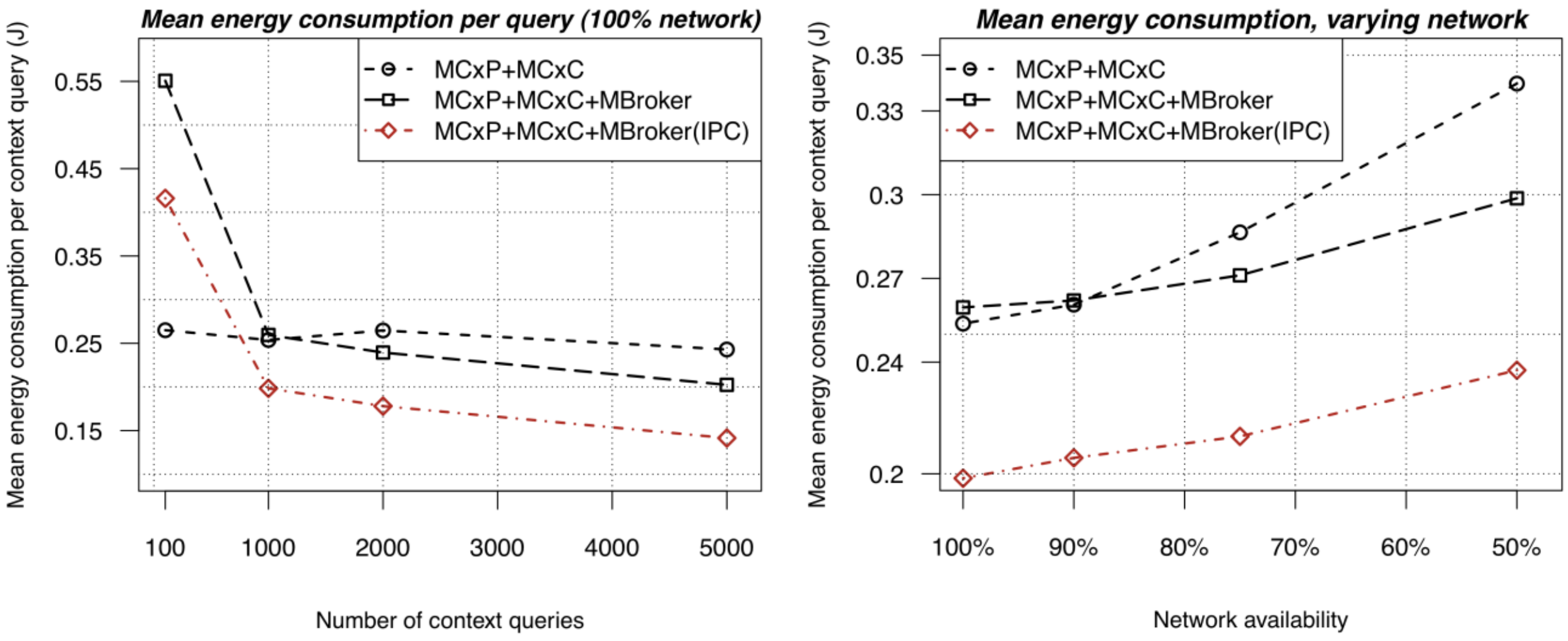

(a) Energy consumption at the device with increasing number of queries during 100% network availability scenario.

(b) Energy consumption at the device during different levels of network availability (1000 context queries).

Figure 16: Extended results of mobile context broker energy consumption experiment with IPC as the communication mechanism between device-based context consumers, providers and the broker.

It is evident that using IPC as the communication mechanism between device-based context consumers, providers and the broker further reduces the mean energy consumption per context query. These energy conservation benefits of the Mobile Context Broker based interaction on the device are pronounced in scenarios with full network availability (Fig. 16a), varying network connectivity conditions (Fig. 16b), and when the broker operates in the bulk mode (Fig. 17). In addition to intrinsic differences in the operating mechanisms that influence the resource utilisation by IPC and HTTP based communication, one of the major factors that influences the overall reduction in energy consumption is that IPC calls take less time to complete on average. Hence the components on mobile devices that utilise IPC spend less *waiting* time and are able to carry out their application life cycles more efficiently.

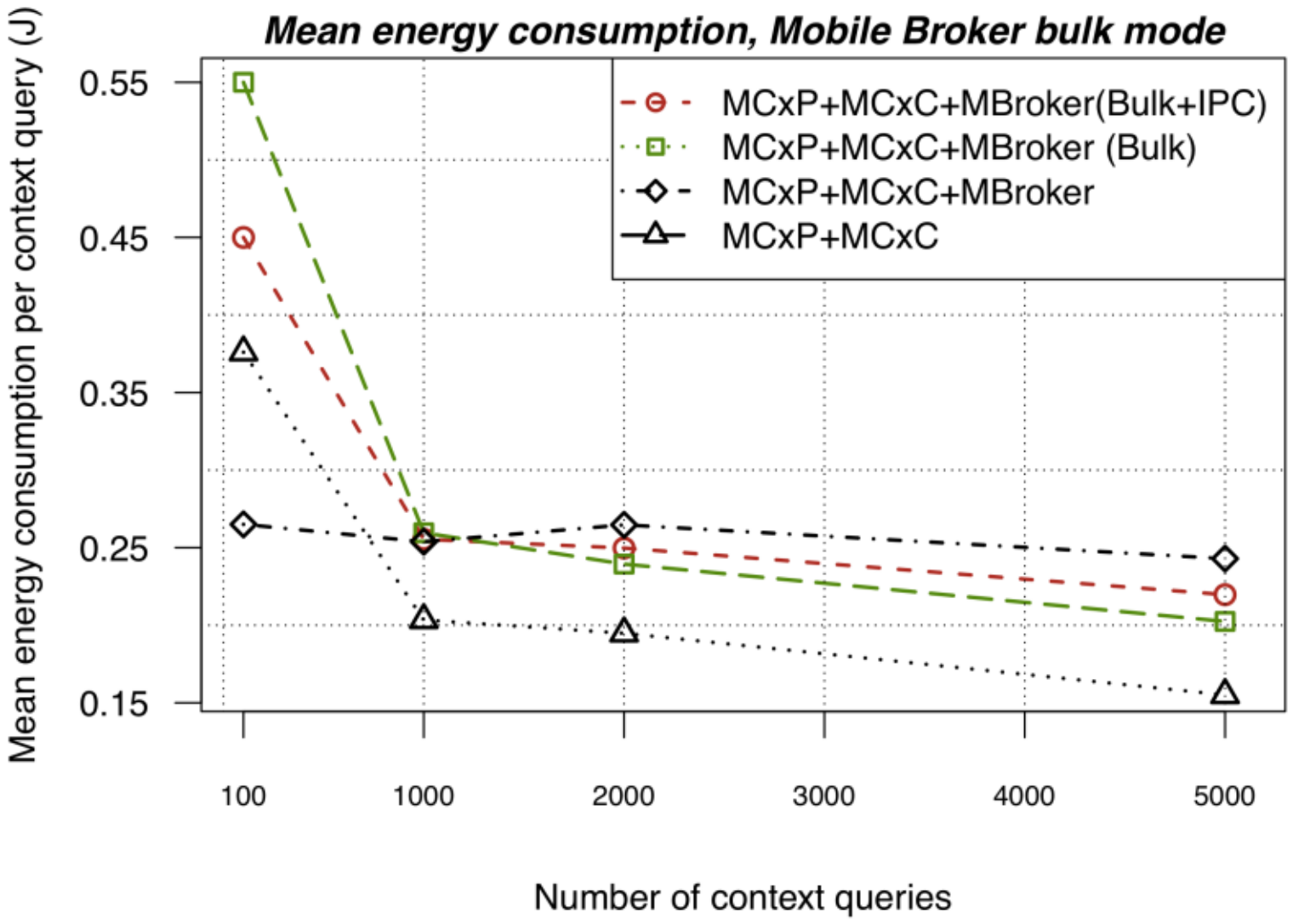

Figure 17: Extended results: Energy consumption at the device with MCxB in bulk mode.

# 7 Conclusion and Future Work

We can conclude from the experimental results discussed in this section that the Mobile Context Broker reduces the overall energy consumption by context consumers/providers during the execution of their context communication functions. The results demonstrate that for longer durations of operations, mean energy consumption per context query is reduced. Furthermore, the Mobile Context Broker also assists in conserving energy during periods of disconnected operation. Though device based context consumers and providers can be developed to limit wasteful execution cycles during network unavailability that incurs a cost in terms of development effort. Based on the results presented in this section, we expect the full utility of the Mobile Context Broker in scenarios where there are a number of context consuming/providing applications and services executing on smart mobile devices and collaborating with cloud-based context services. The evolving technological trends do point towards realisation of such scenarios in the near future. The Mobile Context Broker is a new approach that can accommodate the evolving role of smart mobile devices in context awareness functions while remaining conscious of the critical resource that is the energy reserves available on such devices. The advent of cloud computing and emergence of cloud based service provision of information services will provide new avenues of growth in the types of context-awareness related services that are available to consumers. While the cloud infrastructure provides a number of benefits to the server side infrastructure, it does not directly benefit energy conservation on mobile devices that interact with services hosted in the cloud. The architecture presented in this paper and Mobile Context Broker component is a novel solution for reducing the impact of context-awareness related functions carried out at a mobile device on energy consumption.

The experiments presented in this work are designed to capture a snapshot of a typical use scenario with a limited number of context consumers and providers on the mobile device and in the cloud infrastructure. In the absence of any established benchmarks for evaluation of energy consumption and energy conservation strategies for context-aware service utilisation, this work presents a novel analysis of these issues. The tool utilised in our experiments for recording energy conservation (PowerTutor) has an average error rate of 1% (over 10% second interval) and worst-case error of 2.5% (over a 10 second interval). The accuracy of the reported results, along with the repetition of experiments that establish the significance of communication optimisation and energy consumption on devices with varying hardware capabilities, provides a solid foundation for drawing statistical and logical conclusions from the results. The results have shown that the Mobile Context Broker helps in reducing energy consumption during utilisation of context-awareness related services. Specifically, the mean energy cost per context query and response is reduced. The energy cost of acquiring sensor data and processing context is not considered in these experiments as such costs depend on the type of context and the sensor data acquisition mechanism. This study has focussed on determining only the effect of the Mobile Context Broker on communication and coordination of context between mobile devices i.e. a smartphone and a cloud infrastructure. The factors that aid in improvement in energy consumption include reduction in the amount of network bound communication due to the broker's caching facility and reducing execution cost of consumers and providers during periods of network unavailability. The analysis environment is limited to one mobile device and a strictly defined set of parameters. In future the scope of these experiments will be expanded to include a diverse set of devices, networks and complex query patterns in order to infer more generalised results.

The Mobile Context Broker in the Context Provisioning Architecture presents a novel approach for context communication across mobile devices and network/cloud based services, incorporates the issue of energy conservation in its design and provides the developers of context-aware systems a new approach towards energy-awareness in context-aware systems.